\documentclass[aps,prd,preprint,superscriptaddress,nofootinbib]{revtex4-2}
\usepackage{graphicx}% Include figure files
\usepackage{dcolumn}% Align table columns on decimal point
\usepackage{bm}% bold math
\usepackage{multirow}
\usepackage{color}%
\usepackage[english]{babel}
\usepackage{amsmath,amssymb}
\usepackage{amsfonts}
\usepackage{mathrsfs}
\usepackage{epsfig}
\usepackage{tabularx}
\usepackage{array}
\usepackage{slashed}
\usepackage{breqn}
\usepackage{braket}
\usepackage{mathtools}

\usepackage{floatrow}   
\usepackage{xcolor}
\usepackage{makecell}
\usepackage{hyperref}

\graphicspath{{./figures/}}

\DeclarePairedDelimiter\autobracket{(}{)}
\newcommand{\br}[1]{\autobracket*{#1}}

\newcommand{\cevens}{CE$\nu$NS }
\newcommand{\npe}{n_{\rm PE}}
\newcommand{\pe}{{\rm PE}}

\newcommand{\checked}[2][0.5cm]{%
	\noindent\parbox[t]{#1}{\raggedright\ding{51}}\parbox[t]{\the\dimexpr\linewidth-#1}{#2}%
}

\def\s2tw{{\rm sin ^2 \theta_{W}}}

\makeatletter
\def\slashchar#1{{\mathpalette\c@ncel{#1}}} % TeXbook, bottom of p360
\makeatother

\makeatletter

\begin{document}

%\preprint{}

\title{Constraining Nonminimal Dark Sector Scenarios with the COHERENT Neutrino Scattering Data}

\newcommand{\metu}{Department of Physics, Middle East Technical University, Ankara 06800, Turkey.}
\newcommand{\au}{Department of Engineering Physics, Ankara	University, TR06100 Ankara, Turkey.}

\author{A. Elpe$^{}$\footnote{aelpe@metu.edu.tr}}  \affiliation{ \metu }
\author{E. Akyumuk$^{}$\footnote{esra.akyumuk@metu.edu.tr}}  \affiliation{ \metu }
\author{T.M.~Aliev$^{}$\footnote{taliev@metu.edu.tr}}  \affiliation{ \metu }
\author{L. Selbuz$^{}$\footnote{selbuz@eng.ankara.edu.tr}} \affiliation{ \au }
\author{I.~Turan$^{}$\footnote{ituran@metu.edu.tr}}  \affiliation{ \metu }


\date{\today}

\begin{abstract}
Abelian dark sector scenarios embedded into the two-Higgs doublet model are scrutinized within the Coherent Elastic Neutrino-Nucleus Scattering (CE$\nu$NS) experiment, which was first measured by the COHERENT Collaboration in 2017 with an ongoing effort to improve it since then and recently released   data for the  CsI target in 2022. In the theoretical framework, it is assumed that there is a $U(1)$ gauge group in the dark sector with a nonzero kinetic mixing with the hypercharge field. The COHERENT data for the targets CsI and liquid argon (LAr) are treated in both single and multibin bases to constrain the multidimensional parameter space, spanned by the dark gauge coupling, kinetic mixing parameter and the dark photon mass, of totally seven different representative scenarios which are also compared and contrasted among each other to find out about the most sensitive one to the data. The effect of refined quenching factor is also addressed. 
\end{abstract}

\pacs{XXX}

\maketitle
\newpage

%\section{Introduction}
 
%\tableofcontents
 
\section{Introduction}
\label{intro}
With the discovery of the Standard Model Higgs boson at Large Hadron Collider \cite{CMS:2012qbp,ATLAS:2012yve}, the last missing piece of the Standard Model (SM) was found. While the search for physics beyond the Standard Model at high energies has been continuing, there are low-energy tools which could be used to test physics beyond the SM. Recently measured coherent elastic neutrino-nucleus scattering (\cevens) \cite{COHERENT:2017ipa, COHERENT:2018imc,COHERENT:2020iec,COHERENT:2020ybo} is one of such powerful tools, which has been considered for new physics scenarios \cite{Lindner:2016wff,Billard:2018jnl,AristizabalSierra:2018eqm,Miranda:2019skf}, for performing precision tests of the Standard Model \cite{Miranda:2019skf,Canas:2018rng}, as well as being instrumental for many other implications \cite{Dutta:2019eml,Cadeddu:2019eta,Papoulias:2019lfi,AristizabalSierra:2019ufd,Bellenghi:2019vtc,NUCLEUS:2019igx}.

The \cevens is an elastic  scattering process which takes place between low-energy neutrinos (in the MeV range) and atomic nuclei. It is a weak interaction mediated by $Z$ boson with a tiny momentum transfer (in the keV range), which makes the interaction coherent with the nucleus, whose cross section in the SM is scaled as the square of the number of neutrons in the nucleus ($\sim N^2$). This brings an enhancement to the cross section as compared to the other scatterings in the same energy range.   

The idea of measuring \cevens cross section had been first proposed by Freedman \cite{Freedman:1973yd} and after more than four decades, the COHERENT Collaboration has managed to measure the \cevens by using the targets Cesium-Iodide (CsI[Na]) first \cite{COHERENT:2017ipa, COHERENT:2018imc} and then argon (Ar) \cite{COHERENT:2020iec,COHERENT:2020ybo}. There is also a recent update \cite{COHERENT:2021xmm} by the COHERENT Collaboration with more data on CsI together with various improvements in the analysis like using an updated energy-dependent quenching model, including energy-smearing effect and time dependent efficiency.  There is a very recent analysis \cite{DeRomeri:2022twg} combining full CsI data with the available LAr results which considers various implications like testing the SM, studying the electromagnetic properties of neutrinos (see also \cite{Miranda:2019wdy}) and constraining some new physics scenarios by allowing generalized types of neutrino-neucleus interactions while assuming universal couplings.

It has to be mentioned that all the findings of COHERENT results confirm the  $N^2$ scaling property of the SM. Hence, it becomes one of the suitable testing grounds for physics beyond the SM. \cevens experiments like COHERENT should be considered to be complementary to the search in high-energy collider experiments due to their low-energy coverage, which makes them ideal for searching  axionlike particles \cite{AristizabalSierra:2020rom, Dent:2019ueq}, new fermions \cite{Brdar:2018qqj,Chang:2020jwl}, dark matter searches \cite{Ge:2017mcq,Dutta:2020vop,Dror:2019onn,Dror:2019dib}, neutrino NSI \cite{Giunti:2019xpr,Bernal:2022qba,DeRomeri:2022twg}, and light vector bosons  like the so-called dark photon \cite{deNiverville:2015mwa, Dent:2016wcr, Dutta:2019eml, CONNIE:2019xid, Miranda:2020zji, Cadeddu:2020nbr,Flores:2021kzl} as well as scalar \cite{DeRomeri:2022twg,Flores:2021kzl, Farzan:2018gtr} and tensor particle \cite{DeRomeri:2022twg,Flores:2021kzl} searches.

The discovery of the SM Higgs as the first elementary spin-0 scalar has boosted the interest in extending the scalar sector of the SM, the simplest of which is known as Two Higgs Doublet Models (2HDM). Within the 2HDM scenario it is possible to address both the flavor changing neutral current problem and mass for neutrinos \cite{Campos:2017dgc} by extending the gauged sector of the SM with a $U(1)_D$ group together with right-handed neutrinos added to the spectrum. Various anomaly free models have been constructed and discussed  in high- and low-energy scales \cite{Campos:2017dgc}. The use of \cevens data  as a new physics probe to constrain the 2HDM gauged with $U(1)_D$ is missing in the literature and the current study aimed to provide a thorough investigation on this matter to fill the gap.

The anomaly free extension of the SM (plus right-handed neutrinos) by coupling it with a $U(1)_{B-L}$ gauge group from a hidden sector is a well-established scenario to study dark photon effects through its kinetic mixing with the SM hypercharge gauge field as well as its gauge coupling with the SM fermions. Such a scenario would be naturally a limiting case of the 2HDM mentioned above. All of the viable scenarios will be compared and contrasted with each other in the light of COHERENT data.

In Section \ref{model}, the theoretical framework is summarized and the relevant SM vertices which receive corrections or the new vertices are also listed for completeness. The cross section formulas contributing the \cevens and some basics about the COHERENT data are given in Section \ref{nes} and Section \ref{cohdata}, respectively. After having the statistical method briefly explained, the numerical study is conveyed in Section \ref{analysis} and   we conclude in Section \ref{conc}.

\section{A Nonminimal Dark Sector  Framework}
\label{model}

One way of realizing interactions between the hidden (dark) sector and the visible (SM) one  is through various portals which are dimension-4 operators and hence are free of suppression. A so-called dark vector boson from the hidden sector coupling with the SM is one popular scenario, known as the vector portal. 

In a minimal scenario, such a coupling includes {\it only} a kinetic mixing between the dark vector boson and the weak hypercharge field, which allows the dark vector boson indirect access to the SM fermions. If the vector boson of the dark sector is a gauge field of a dark group, say $U(1)_D$ (again in the minimalistic approach), having the SM fermions charged under  $U(1)_D$ would allow them to get a direct coupling with the dark vector boson. This will require checking additional gauge-anomaly conditions which restrict the dark quantum charges of the SM fermions. $B-L$ is one popular choice for quantum charges which makes the overall scenario anomaly free when the SM is extended with the right-handed neutrinos.

In the above framework, both the visible and dark sectors are assumed to be minimal for the sake of taking advantage of mainly the predictive power of the scenario. Either or both sectors could be enlarged, when the SM being an effective theory is especially considered, which leads to nonminimal scenarios.  

One popular way of extending the SM while keeping the dark sector minimal (specifically, assuming an Abelian dark-gauge group $U(1)_D$) is through its scalar part, simply by adding another $SU(2)$ doublet\footnote{In the SM, considering one SU(2) scalar doublet is the simplest choice but not a theoretical requirement.}, known as the two-Higgs doublet models (2HDM) \cite{Lee:1973iz} which have motivations from supersymmetry \cite{Haber:1984rc}, axions \cite{Kim:1986ax}, baryogenesis \cite{Trodden:1998qg,Gleiser:1992ch,Joyce:1994zt,Funakubo:1993jg} etc.  On the other hand, adding right-handed neutrinos to the particle content of the construct would not only address the issue of neutrino masses but also usually needed for satisfying anomaly equations due to gauging the 2HDM under the additional dark group $U(1)_D$. The details of the scenarios have been worked out in other studies \cite{Lee:2013fda,Campos:2017dgc,Lindner:2018kjo} (see also \cite{Fayet:1989mq,Fayet:1990wx,Fayet:1986rh,Fayet:2016nyc,Fayet:2020bmb} for earlier theoretical studies including different aspects of the mixing among the gauge bosons) and here we only reproduce the relevant part of the model whose Lagrangian terms are given below
\begin{eqnarray}
	{\cal L} &=& {\cal L}_{\rm Gauge}^{\rm KE} + {\cal L}_{\rm Gauge}^{\rm KM+Mass} + {\cal L}_{\rm Scalar}^{\rm KE}+ {\cal L}_{\rm Fermion}^{\rm KE}+\dots\,, \nonumber\\
	{\cal L}_{\rm Gauge}^{\rm KE} &=& -\frac14 W_{3\mu\nu} W_3^{\mu\nu}-\frac14 Y_{\mu\nu} Y^{\mu\nu}-\frac14 X^0_{\mu\nu}
	X^{0\mu\nu} + \dots\,,\\
	{\cal L}_{\rm Gauge}^{\rm KM+Mass} &=& -\frac12 \sin\!\epsilon\ X^0_{\mu\nu}Y^{\mu\nu} +  \frac12 m_{X}^2 X^0_\mu X^{0\mu}\,,\label{LGaugeKM} \\
	{\cal L}_{\rm Scalar}^{\rm KE} &=& (D_\mu \phi_1)^\dagger(D^\mu \phi_1) + (D_\mu \phi_2)^\dagger(D^\mu \phi_2)\,,\label{LGaugeSKE}\\
%	&=&\frac12 m_{Z_0}^2 \widetilde{W}_{3\mu} \widetilde{W}_3^\mu +(\dots)X_\mu \widetilde{W}_3^\mu+\dots
	 %	\label{LGaugeSKE}\\	
	{\cal L}_{\rm Fermion}^{\rm KE} &=& \sum_i \bar{f}_i i{\slashchar D} f_i\,.
	\label{LGauge}
\end{eqnarray}
Here $X^0_\mu$ is the gauge field of $U(1)_D$ while $Y_\mu$ and $W_{3\mu}$ are the usual weak hypercharge field of $U(1)_Y$ and the third of weak gauge fields, respectively. %$X_\mu$ and $\widetilde{W}_{3\mu}$ are the rotated (but not yet physical) fields which will be defined later in this section.  
$\phi_1$ and $\phi_2$ in 	${\cal L}_{\rm Scalar}^{\rm KE}$ are the usual scalar doublets under $SU(2)_L$. $\sin\epsilon$ is the strength of the kinetic mixing  between $U(1)_Y$ and $U(1)_D$. 
%$m_{Z_0}$ would be the mass of the electroweak gauge boson $Z^0$ in the SM limit of the current scenario. 
$m_X$ represents the Stueckelberg mass parameter for the dark gauge field $X^0_\mu$, which will be explained briefly.

Unlike in the non-Abelian case, the mass generation for the Abelian gauge bosons does not necessarily require the existence of a Higgs mechanism since there is no issue of unitarity or renormalizability for Abelian gauge theories. Hence alternative ways of mass generation  other than the Higgs mechanism can be employed for such sectors like $U(1)_D$ while in the SM part, masses are acquired through the Higgs mechanism. One of the popular ways is the so-called Stueckelberg mechanism \cite{Stueckelberg:1938hvi} where the $U(1)$ gauge boson couples with the derivative of an axionic (dark) scalar field in a gauge invariant manner. We assume that the scalar field from the dark sector is charged under $U(1)_D$ only so that the $m_X$ mass term in Eqn.~(\ref{LGaugeKM}) is obtained. The Stueckelberg extension of the SM with \cite{Feldman:2007wj} and without the kinetic mixing \cite{Kors:2004dx} have been discussed in detail. The two-Higgs doublet model extended with an additional $U(1)$ gauge symmetry has also been studied in \cite{Panotopoulos:2011xb} by considering the Stueckelberg contribution to the mass terms.

The Lagrangian ${\cal L}_{\rm Fermion}$ contains the kinetic energies and electroweak interactions of the fermions. At this stage, the relevant part of the  covariant derivative in the  $(Y_\mu,W_{3\mu},X^0_\mu)$ basis would look like
\begin{eqnarray}
	D_\mu =\partial_\mu-ig_W\, t^3\, W_{3\mu} - i g_Y Y Y_\mu\nonumber -i g_D  Q_D^\prime X^0_\mu + \dots
\end{eqnarray}
where $t^3$,  $Y$, and $Q_D^\prime$ are the $SU(2)_L$ generator, weak hypercharge and dark charges of the field that the covariant derivative is acting on, respectively. Here $g_W$, $g_Y$ and $g_D$ are the corresponding gauge coupling constants. 

\begin{table*}[h]
	\caption{Dark quantum charges of the fields under $U(1)_D$, adapted from Ref. \cite{Campos:2017dgc}. Note that there is a difference in the convention to define the covariant derivative where we use the Peskin and Schroeder convention.} \label{ChargeTable}
	\addtolength{\tabcolsep}{0.7pt}
	\begin{tabularx}{1\textwidth}{@{\extracolsep{\fill}} lrrrrrrrr}
		\hline \hline \qquad Fields & $u_R$ & $d_R$ & $Q_L$ & $L_L$ & $e_R$ & $\nu_R$ &
		$\phi_2$ & $\phi_1$     
		\\\hline \\ [-3.1ex]
		$\;$  Dark Charges & $Q_u'$ & $Q_d'$ &$\frac{Q_u'+Q_d'}{2}$ & $-\frac{3(Q_u'+Q_d')}{2}$ & $-(2Q_u'+Q_d')$ & $-(Q_u'+2Q_d')$ & $\frac{Q_u'-Q_d'}{2}$ & $\frac{5Q_u'+7Q_d'}{2}$ \\[0.2em]\hline
		$\;$  Model C&$\frac{1}{4}$ &$-\frac{1}{2}$ &$-\frac{1}{8}$&$\frac{3}{8}$&0&$\frac{3}{4}$&$\frac{3}{8}$&$-\frac{9}{8}$ \\
		$\;$ Model D &$\frac{1}{2}$&0&$\frac{1}{4}$&$-\frac{3}{4}$&$-1$&$-\frac{1}{2}$& $\frac{1}{4}$&$\frac{5}{4}$    \\
		$\;$  Model E&0&$\frac{1}{2}$&$\frac{1}{4}$&$-\frac{3}{4}$&$-\frac{1}{2}$&$-1$&$-\frac{1}{4}$&$\frac{7}{4}$ \\
		$\;$  Model F &$\frac{2}{3}$&$\frac{1}{3}$&$\frac{1}{2}$&$-\frac{3}{2}$&$-\frac53$&$-\frac{4}{3}$&$\frac{1}{6}$&$\frac{17}{6}$ \\
		$\;$ Model G &$-\frac{1}{6}$&$\frac{1}{3}$&$\frac{1}{12}$&$-\frac{1}{4}$&0&$-\frac{1}{2}$&$-\frac{1}{4}$&$\frac{3}{4}$    \\
		$\;$  Model $B-L$&$\frac{1}{6}$&$\frac{1}{6}$&$\frac{1}{6}$&$-\frac{1}{2}$&$-\frac{1}{2}$& $-\frac{1}{2}$&0&1\\\hline
		$\;$  Minimal $B-L$&$\frac{1}{6}$&$\frac{1}{6}$&$\frac{1}{6}$&$-\frac{1}{2}$&$-\frac{1}{2}$& $-\frac{1}{2}$&0& $-$\\
		\hline\hline
	\end{tabularx}
\end{table*}

The details of the scalar sector as well as the discussion of anomaly cancellations of the model are given in \cite{Campos:2017dgc}. Some of the relevant parts are going to be repeated here. For example, the dark charge assignment of the SM fields satisfying the anomaly conditions are listed in Table~\ref{ChargeTable}. The mass terms for the gauge fields that will originate from Eqn.~(\ref{LGaugeSKE})  are not diagonal due to the kinetic mixing in Eqn.~(\ref{LGaugeKM})  as well as the kinetic energy of the scalar doublets in Eqn.~(\ref{LGaugeSKE}) when they take their vacuum expectation values (VEVs), that is, 
\begin{eqnarray}
	\displaystyle
	\langle \phi_i \rangle = \left(
	\begin{array}{c}
	    0 \\
	    \frac{v_i}{\sqrt{2}}	
		\end{array}
		\right)\,\quad i=1,2\,.
\end{eqnarray}

%\begin{table}[htb]
%	\begin{center}
%		\begin{tabularx}{1.5\textwidth} { 
%				>{\raggedright\arraybackslash}X 
%				>{\centering\arraybackslash}X 
%				>{\centering\arraybackslash}X 
%				>{\centering\arraybackslash}X 
%				>{\centering\arraybackslash}X 
%				>{\centering\arraybackslash}X 
%				>{\centering\arraybackslash}X 
%				>{\raggedleft\arraybackslash}X 
%				>{\raggedleft\arraybackslash}X 
%			}%{ccccccccc} 
%			\hline\hline
%			Fields & $u_R$ & $d_R$ & $Q_L$ & $L_L$ & $e_R$ & $\nu_R$ & $\phi_2$ & $\phi_1$ \\
%			\hline
%			Dark Charges & $Q_u'$ & $Q_d'$ & $\frac{Q_u'+Q_d'}{2}$ & $-\frac{3(Q_u'+Q_d')}{2}$ & $-(2 Q_u'+Q_d')$ & $-(Q_u'+ 2 Q_d')$  & $\frac{Q_u'-Q_d'}{2}$ & $\frac{5Q_u'}{2} + \frac{7Q_d'}{2}$\\
%			$\mU(1)_C$ & $\frac14$ & $-\frac12$ & -$\frac18$ & $\frac38$ & 0 & $\frac34$ & $\frac38$ & $\frac98$ \\			
%			$\mU(1)_D$ & $\frac12$ & 0 & $\frac14$ & $-\frac34$ & -1 & $-\frac12$ & $\frac14$ & $\frac54$ \\			
%			$\mU(1)_E$ & $0$ & $\frac12$ & $\frac14$ & $-\frac34$ & $-1$ & $\frac74$ & $-\frac14$ \\			
%			$\mU(1)_F$ & $\frac23$ & $\frac13$ & $\frac12$ & $-\frac32$ & -2 & $-\frac43$ & $\frac16$ & $\frac{17}{6}$ \\			
%			$\mU(1)_G$ & $-\frac16$ & $\frac13$ & $\frac{1}{12}$ & $-\frac14$ & 0 & $-\frac12$ & $-\frac14$ & $-\frac34$ \\
%			$\mU(1)_{BL}$ & $\frac16$ & $\frac16$ & $\frac16$ & $-\frac12$ & $-\frac12$ & $-\frac12$ & 0 & 1 \\
%			\hline\hline
%		\end{tabularx}
%		\caption{Dark quantum charges of the fields under $U(1)_D$, taken from \cite{Campos:2017dgc}.}
%		\label{ChargeTable}
%	\end{center}
%\end{table}

Having said that the original gauge basis in the neutral sector, i.e., $(Y_\mu,W_{3\mu},X^0_\mu)$,  is not diagonal, the physical basis, say  $(A_\mu, Z_\mu, A'_\mu)$, can be obtained by making three successive rotations. The first two of these are the ones to eliminate the kinetic mixing term and the usual Weinberg angle rotations, which results in the rotated gauge basis  $(A_\mu, \widetilde{W}_{3\mu}, X_\mu)$, given by in terms of $(Y_\mu,W_{3\mu},X^0_\mu)$,

\begin{eqnarray}
	\left(
	\begin{array}{c}
		A_\mu \\[0.3em]
		\widetilde{W}_{3\mu}\\[0.3em]
		X_\mu
	\end{array}
	\right) &=& 
	\left(
	\begin{array}{ccc}
		\cos\theta_W & \sin\theta_W & \sin\epsilon \cos\theta_W \\[0.3em]
		-\sin\theta_W & \cos\theta_W & -\sin\epsilon \sin\theta_W \\[0.3em]
		0 & 0 &\cos\epsilon
	\end{array}
	\right)
	\left(
	\begin{array}{c}
		Y_\mu \\[0.3em]
		W_{3\mu}\\[0.3em]
		X^0_\mu
	\end{array}
	\right)\,.
\end{eqnarray}
At this stage the mass term becomes 
\begin{align*}
	{\cal L}_{\rm Gauge}^{\rm KM+Mass} + {\cal L}_{\rm Scalar}^{\rm KE} \supset\, & \frac{1}{2} m_X^2 \sec^2\!\epsilon\ X_\mu X^\mu + \sum_{i=1}^2 \br{ D_\mu \left < \phi_i \right > }^\dagger \br{ D_\mu \left < \phi_i \right > }\nonumber\\
	=&\frac{1}{2}\begin{pmatrix}
		A_\mu & \widetilde{W}_{3\mu} & X_\mu
	\end{pmatrix}
	\left(
\begin{array}{ccc}
	0 & 0 & 0 \\[0.3em]
	0 & a & b \\[0.3em]
	0 & b & c
\end{array}
\right)
	 \begin{pmatrix}
		A^\mu \\ \widetilde{W}_{3}^\mu \\ X^\mu
	\end{pmatrix}.
\end{align*}
where the parameters $a,b,$ and $c$ are given 
\begin{eqnarray}
	a &=& m_{Z_0}^2, \nonumber\\
	b &=& m_{Z_0}^2\left(\cos^2\!\beta\ \Delta_1 + \sin^2\!\beta\ \Delta_2\right)\,,\\
	%&=& m_{W_3}^2 \sin\!\theta_W \tan\!\epsilon  + g_D\, v\, m_{W_3} \sec\!\epsilon\, \br{\cos^2\!\beta\ Q_D^{\phi_1} + \sin^2\!\beta\ Q_D^{\phi_2}},\nonumber\\
	c &= &  m_X^2 \sec^2\!\epsilon  + m_{Z_0}^2 \left(\cos^2\!\beta\ \Delta_1^2 + \sin^2\!\beta\ \Delta_2^2\right)\,. \nonumber
	%\\	&=&  m_X^2 \sec^2\!\epsilon  + m_{W_3}^2 \sin^2\!\theta_W \tan^2\!\epsilon   - 2\,g_D\, v\, m_{W_3}\, \sin\!\theta_W \tan\!\epsilon\; \sec\!\epsilon \,  \br{\cos^2\!\beta\ Q_D^{\phi_1} + \sin^2\!\beta\ Q_D^{\phi_2}}\nonumber\\ 
%	&&+ g_D^2\ v^2 \sec^2\!\epsilon \left[\cos^2\!\beta\ \left(Q_D^{\phi_1}\right)^{\!2} + \sin^2\!\beta\ \left(Q_D^{\phi_2}\right)^{\!2}\right]\,.
\end{eqnarray}
Here the functions $\Delta_i$ are defined as follows
\begin{eqnarray}
	\Delta_i = \sin\!\theta_W \tan\!\epsilon - \frac{g_D v}{m_{Z_0}}\sec\!\epsilon\, Q_D^{\phi_i}\,,\qquad i=1,2,
	\end{eqnarray}
%\begin{eqnarray}
%	a &=& m_{W_3}^2, \nonumber\\
%	b %&=& \cos^2\!\beta\ \widetilde{Q}_D^{\phi_1} + \sin^2\!\beta\ \widetilde{Q}_D^{\phi_2}\\
%	&=& m_{W_3}^2 \sin\!\theta_W \tan\!\epsilon  + g_D\, v\, m_{W_3} \sec\!\epsilon\, \br{\cos^2\!\beta\ Q_D^{\phi_1} + \sin^2\!\beta\ Q_D^{\phi_2}},\nonumber\\
%	c %&= &  m_X^2 \sec^2\!\epsilon  + m_{W_3}^2 \left[\cos^2\!\beta\ \left(\widetilde{Q}_D^{\phi_1}\right)^{\!2} + \sin^2\!\beta\ \left(\widetilde{Q}_D^{\phi_2}\right)^{\!2}\right] \nonumber\\
%		&=&  m_X^2 \sec^2\!\epsilon  + m_{W_3}^2 \sin^2\!\theta_W \tan^2\!\epsilon   - 2\,g_D\, v\, m_{W_3}\, \sin\!\theta_W \tan\!\epsilon\; \sec\!\epsilon \,  \br{\cos^2\!\beta\ Q_D^{\phi_1} + \sin^2\!\beta\ Q_D^{\phi_2}}\nonumber\\ 
%	&&+ g_D^2\ v^2 \sec^2\!\epsilon \left[\cos^2\!\beta\ \left(Q_D^{\phi_1}\right)^{\!2} + \sin^2\!\beta\ \left(Q_D^{\phi_2}\right)^{\!2}\right]\,.
%\end{eqnarray}
and $\tan\beta\equiv v_2/v_1$ (with $\sqrt{v_1^2+v_2^2} = v = 246\, \textrm{GeV}$) is the ratio of the VEVs, usually defined in the 2HDMs. $Q_D^{\phi_i}$ is the dark charge of the scalar doublet $\phi_i\,,\ i=1,2$. $m_{Z_0}=\frac12 \sqrt{g_Y^2+g_W^2}\,v$ is the mass of the SM $Z$ boson.

The third and last rotation is needed due to $b\ne 0$ and the rotation angle $\xi$ should satisfy 	$\tan 2\xi = 2b/(a-c)$ and the final form of the overall rotation matrix from the $(Y_\mu,W_{3\mu},X^0_\mu)$ basis to $(A_\mu,Z_\mu,A'_\mu)$ becomes
\begin{eqnarray}
	\!\!\!\!\!\!\left(
	\begin{array}{c}
		A_\mu \\[0.3em]
		Z_\mu\\[0.3em]
		A^\prime_\mu
	\end{array}
	\right) &=& 
	\left(
	\begin{array}{c@{\extracolsep{0.5cm}}c@{\extracolsep{0.6cm}}c}
		\cos\theta_W & \sin\theta_W &  \sin\!\epsilon \cos\theta_W\\[0.3em]
		-\cos\xi\sin\theta_W & \cos\xi \cos\theta_W & \sin\xi \cos\!\epsilon -\cos\xi\sin\!\epsilon \sin\theta_W  \\[0.3em]
		\sin\xi \sin\theta_W  & -\sin\xi \cos\theta_W &\cos\xi \cos\!\epsilon +\sin\xi\sin\!\epsilon \sin\theta_W
	\end{array}
	\right)
	\left(
	\begin{array}{c}
		Y_\mu \\[0.3em]
		W_{3\mu}\\[0.3em]
		X^0_\mu
	\end{array}
	\right)\label{fullmixing}
\end{eqnarray}
The corresponding eigenvalues are 
\begin{eqnarray}
	M_A^2 &=& 0, \\
	M_{A^{\prime}}^{2} &=& \ m_X^2 \cos^2\!\xi \sec^2\!\epsilon + \frac14 g_D^2 v^2 \cos^2\!\xi
	\sec^2\!\epsilon \left[\cos^2\!\beta \left( Q_D^{\phi_1 } \right)^2+ \sin^2\!\beta
	\left( Q_D^{\phi_2 }\right)^2 \right] \nonumber \\	
	&& + g_D\ v\ m_{Z_0}\cos\xi \sec\! \epsilon\  	\left(\cos^2\!\beta\
	Q_D^{\phi_1} + \sin^2\!\beta\
	Q_D^{\phi_2}\right)(\sin\xi - \cos\xi
	\sin\theta_W \tan\!\epsilon) \nonumber \\
	&& +m_{Z_0}^2(\sin\xi -\cos\xi \sin \theta_W \tan\!\epsilon)^2, \label{eqn:MAp}\\
	M_{Z}^{2} &=& 	m_X^2 \sin^2\!\xi \sec^2\!\epsilon 
+\frac14 	g_D^2 v^2 \sin^2\!\xi \sec^2\!\epsilon 
 \left[\cos^2\!\beta
	\left( Q_D^{ \phi_1 }\right)^2+\sin^2\!\beta
	\left( Q_D^{ \phi_2 }\right)^2 \right] \nonumber \\
	&& -g_D\ v\ m_{Z_0}\sin\xi\, \sec\!\epsilon 
\left(\cos^2\!\beta\
Q_D^{\phi_1} + \sin^2\!\beta\
Q_D^{\phi_2}\right)(\cos \xi + \sin \xi \sin\!\theta_W  \tan\!\epsilon) \nonumber \\
	&& +m_{Z_0}^2(\cos\xi +\sin\xi \sin \theta_W \tan\!\epsilon)^2.\label{eqn:MZ}
\end{eqnarray}
These are the mass squares of the photon ($A_\mu$), dark photon ($A'_\mu$) and electroweak neutral boson ($Z_\mu$), respectively. The mass-squared expressions for the gauge bosons in the minimal $B-L$ model can be read off from Eqns.~(\ref{eqn:MAp})  and ~(\ref{eqn:MZ}) by taking the limit $g_D\to 0$.

The SM $Z$ boson mass, $M_Z$, receives corrections due to being mixed with the initial $U(1)_D$ gauge field, $X^0_\mu$, and coupling with the Higgs doublets having nonzero dark charges. Even though the analytical expression of $M_Z$ in Eqn.~(\ref{eqn:MZ}) differs from the SM mass, $m_{Z_0}=\frac12 \sqrt{g_Y^2+g_W^2}\,v$\,, it is observed that the numerical value of $M_Z$ lies on the SM prediction for almost all the parameter space as long as $m_X$ is not greater than $\sim \cal O (\rm GeV)$ together with small enough kinetic mixing parameter $\sin\epsilon$.

\begin{figure*}[htb]
	$\begin{array}{cc}
		\vspace*{0.3cm}
		\hspace*{-0.2cm}
		\includegraphics[scale = 0.105]{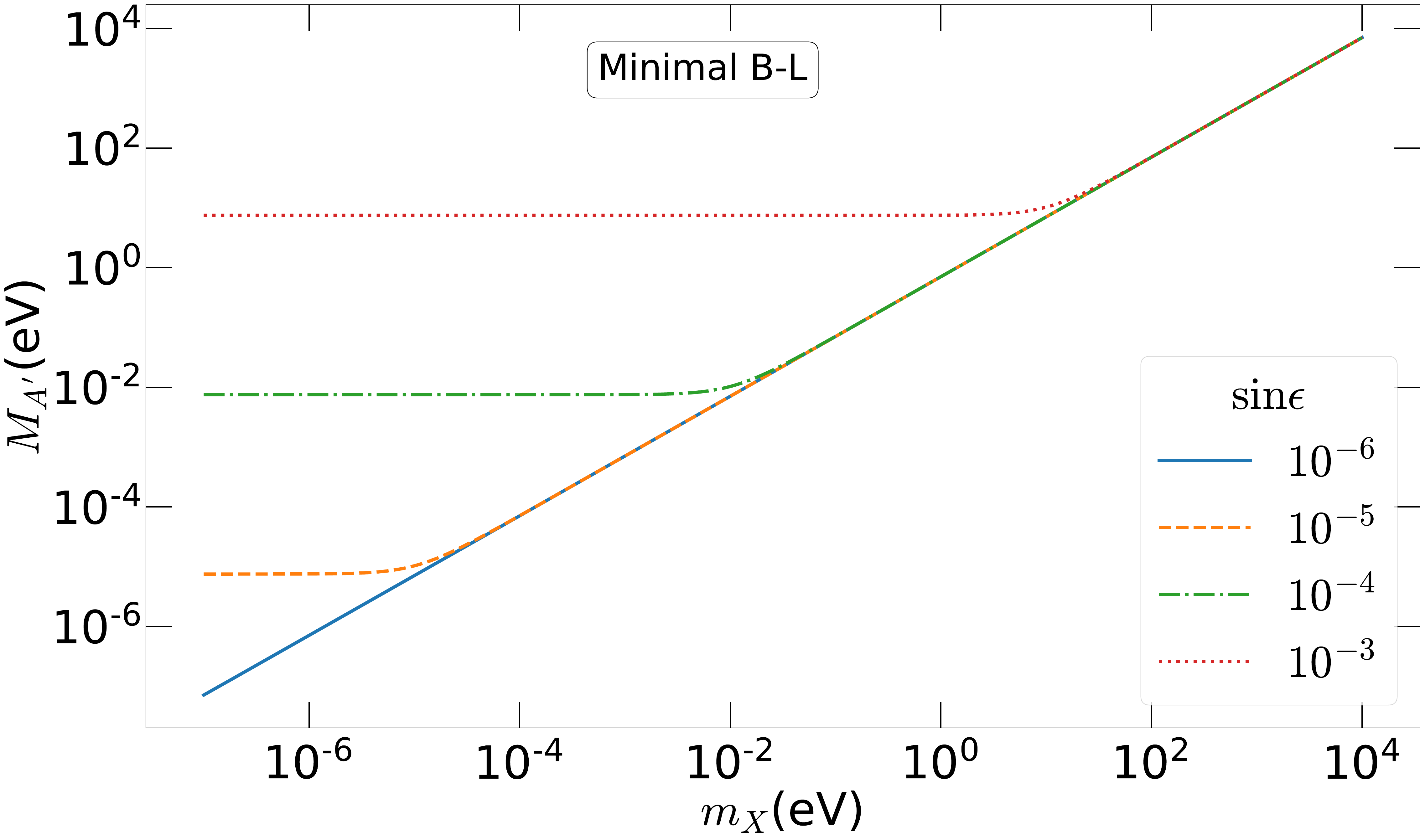} &
		\includegraphics[scale = 0.105]{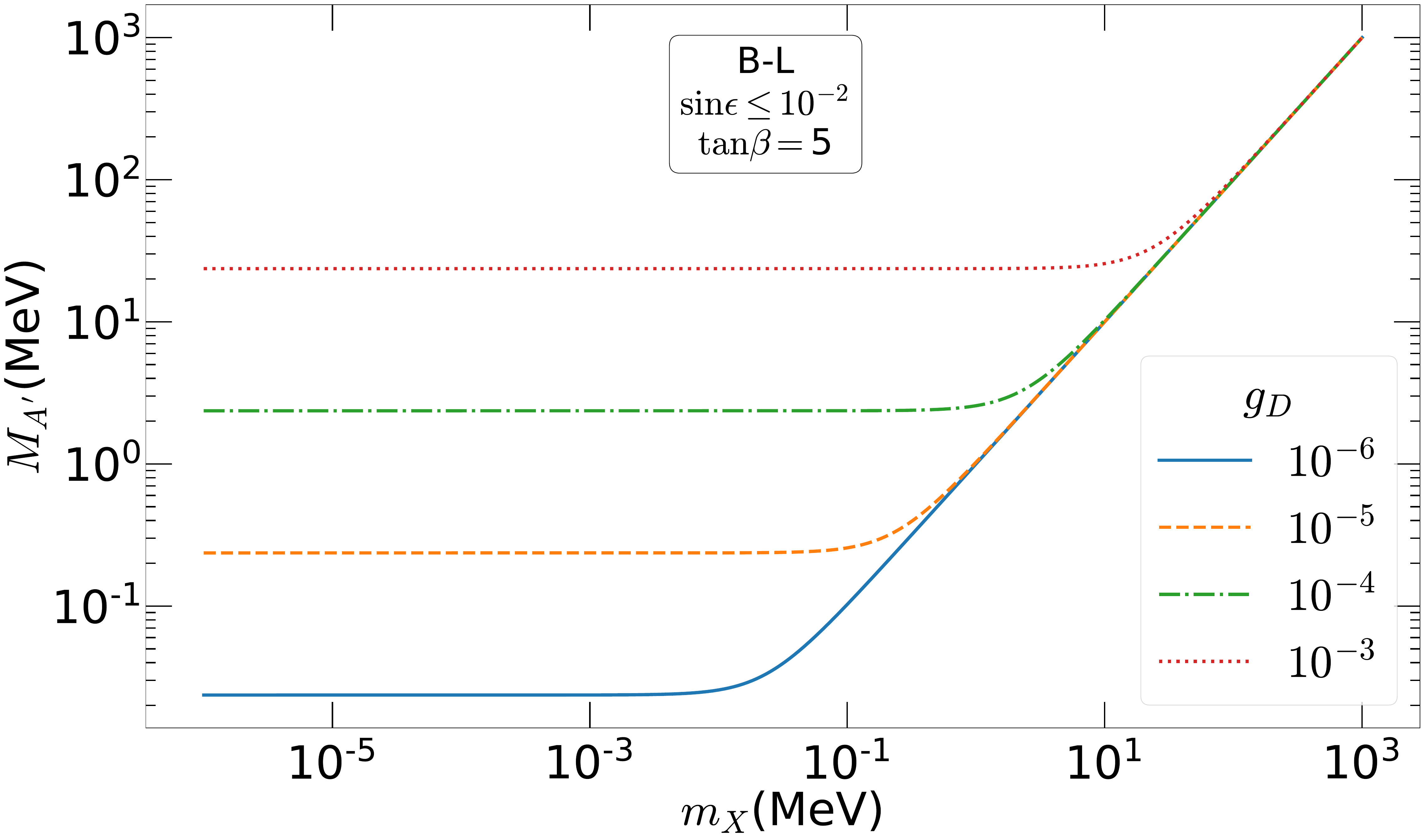}
	\end{array}$
	\vskip -0.5cm
	\caption{The mass of the dark photon, $M_{A'}$, as a function of $m_X$ for various $\sin\epsilon$ in the minimal $B-L$ model (left) and for various $g_D$ in the Two-Higgs Doublet $B-L$ model (right).}
	\label{fig:comparemass}
\end{figure*}
\begin{table*}[htb]
	%	\captionsetup{justification=justified,indention=-2.5cm}
	\caption{The relevant vertex factors contributing to the CE$\nu$NS in the two-Higgs doublet models extended with a dark $U(1)_D$ group. A shorthand notation is used for the trigonometric expressions. For example, $(s_\xi\,, t_\epsilon)$ stand for $(\sin \xi\,, \tan\!\epsilon$) and similar for the others.}
	\label{tab:vertex}
	%\begin{tblr}{Q[c,h]X[c]X[c]}	
	\begin{tabularx}{0.99\linewidth}{X c >{\centering\arraybackslash}p{1.2cm}>{\centering\arraybackslash}c >{\centering\arraybackslash}c }
		\hline\hline
		Vertices &   $C_V^{f}$ & &  $C_A^{f}$  \\
		\hline\\[-1.5em]
		\includegraphics[scale=0.8]{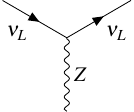} 
		&
		\makecell[l]{\vspace*{1.5cm}
			{ $-\frac{g_D s_\xi  \br{7 Q'_d + 5 Q'_u} }{8c_\epsilon} 
			+\frac{e \br{c_\xi + s_\xi t_\epsilon s_W}}{4 c_W s_W}$}}
		& &
				\makecell[l]{\vspace*{1.5cm}
			{ $ -\frac{g_D s_\xi  \br{ Q'_d -  Q'_u} }{8c_\epsilon}
			-\frac{ e \br{c_\xi + s_\xi t_\epsilon s_W} }{4 c_W s_W}$}}\\[-10mm]
		%\hline
		\includegraphics[scale=0.8]{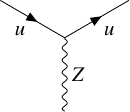} 
		&
				\makecell[l]{\vspace*{1.5cm}
			{$ \frac{ g_D s_\xi  \br{ Q'_d + 3 Q'_u} }{8 c_\epsilon}
			+\frac{ e c_\xi  \br{8s_W^2-3}+ 5e s_\xi t_\epsilon s_W}{12 c_W s_W}$
		}}
		& &
				\makecell[l]{\vspace*{1.5cm}
			{$ -\frac{g_D s_\xi  \br{ Q'_d -  Q'_u}}{8c_\epsilon}
			-\frac{ e  \br{c_\xi + s_\xi t_\epsilon s_W}}{4 c_W s_W}$
		}}\\[-10mm]
		%\hline
		\includegraphics[scale=0.8]{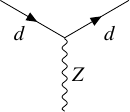} 
		&
				\makecell[l]{\vspace*{1.5cm}
			{$ \frac{g_D s_\xi  \br{ 3 Q'_d + Q'_u} }{8c_\epsilon}
			+\frac{e c_\xi  \br{4s_W^2-3} +  e s_\xi t_\epsilon s_W}{12 c_W s_W}$}}
		& &
				\makecell[l]{\vspace*{1.5cm}
			{$ \frac{g_D s_\xi \br{ Q'_d -  Q'_u} }{8c_\epsilon } 
			+\frac{  e  \br{c_\xi + s_\xi t_\epsilon s_W}}{4 c_W s_W} $
		}} \\[-8mm]
		\hline
		&   $C_V^{'f}$ & &  $C_A^{'f}$ \\
		\hline\\[-6mm]
		\includegraphics[scale=0.8]{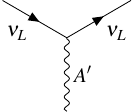} 
		&
		\makecell[l]{\vspace*{1.5cm}{$ \frac{g_D c_\xi \br{7 Q'_d + 5 Q'_u} }{8c_\epsilon } 
			+\frac{  e  \br{s_\xi - c_\xi t_\epsilon s_W} }{4 c_W s_W} $
		}}
		& &
		\makecell[l]{\vspace*{1.5cm}{$ -\frac{g_D c_\xi \br{Q'_d - Q'_u}}{8c_\epsilon } 
			+\frac{  e  \br{s_\xi - c_\xi t_\epsilon s_W} }{4 c_W s_W} $
		}}\\[-10mm]
		%\hline
		\includegraphics[scale=0.8]{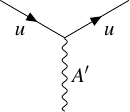} 
		&
		\makecell[l]{\vspace*{1.5cm}{$ \frac{ g_D c_\xi  \br{ Q'_d + 3 Q'_u}}{8c_\epsilon }
			+ \frac{ e  s_\xi  \br{8s_W^2-3}- 5e c_\xi t_\epsilon s_W}{12 c_W s_W}$
		}}
		& &
		\makecell[l]{\vspace*{1.5cm}{$ -\frac{g_D c_\xi  \br{ Q'_d -  Q'_u} }{8c_\epsilon }
			+\frac{ e  \br{s_\xi - c_\xi t_\epsilon s_W}}{4 c_W s_W}$
		}}\\[-10mm]
		%\hline
		\includegraphics[scale=0.8]{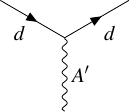} 
		&
		\makecell[l]{\vspace*{1.5cm}{$ \frac{ g_D c_\xi \br{ 3 Q'_d + Q'_u}}{8 c_\epsilon } 
			-\frac{ e  s_\xi  \br{4s_W^2-3} - e c_\xi t_\epsilon s_W}{12 c_W s_W} $
		}}
		& &
		\makecell[l]{\vspace*{1.5cm}{$   \frac{g_D c_\xi  \br{ Q'_d -  Q'_u}}{8c_\epsilon }
			- \frac{ e  \br{s_\xi - c_\xi t_\epsilon s_W}}{4 c_W s_W}$
		}}\\[-8mm]
		\hline\hline
	\end{tabularx}	
	%	\end{tblr}
\end{table*}

In Fig.~\ref{fig:comparemass}, the mass of the dark photon, $M_{A'}$, as a function of the Stueckelberg mass parameter, $m_X$, for various $\sin\!\epsilon$ values in the minimal $B-L$ model (the left panel) and for various $g_D$ values in the two-Higgs doublet $B-L$ model (the right panel) is depicted. The two-Higgs doublet $B-L$ model is chosen as a representative scenario for the ones listed in Table~\ref{ChargeTable}. In the minimal $B-L$ model, $M_{A'}$ has shown sensitivity to the kinetic mixing parameter and as $m_X$ gets smaller, $m_{Z_0}$ term in the mass expression starts to dominate so that we get  different flat curves for different $\sin\!\epsilon$ values. In the two-Higgs doublet $B-L$ model (nonminimal $B-L$ model), $M_{A'}$ has contributions from the Higgs doublets (due to their nonzero dark charges), proportional to the gauge coupling $g_D$ and it becomes a critical parameter. As seen from Fig.~\ref{fig:comparemass}, when $m_X$ becomes smaller than a critical value ($m_X^{\rm critical}$), the dark photon mass is only determined by the terms proportional to $g_D$. Therefore, for a fixed $\sin\epsilon$ or $g_D$ in the region $m_X\le m_X^{\rm critical}$, there will be a nonzero minimum value for the dark photon mass. This feature will be useful in the numerical discussion section.  

%\begin{equation}
%	\begin{aligned}
%        M_A^2& = 0, \\
%		M_{A^{\prime}}^{2}& = m_X^2\cos^{2}\xi\sec^{2}\epsilon + m_{Z_0}^{2}\left[\sin\xi-\cos\xi\sin\theta_{W}\tan\epsilon\right]^{2},\\
%		M_{Z}^{2}& =  m_X^2\sin^{2}\xi\sec^2\epsilon +  m_{Z_0}^2\left[\cos\xi + \sin\xi\sin\theta_{W}\tan\epsilon\right]^2.
%	\end{aligned}
%\end{equation}

%\subsection{Vertex Factors}
%\label{vertices}
%-------VERTEX FACTORS-----------------------
%--------Z---------

At the end of the section, we prefer to give the list of vertex factors relevant to the \cevens process. They are all listed in Table~\ref{tab:vertex}, based on the parametrization given in the following Lagrangian terms,
$$%\begin{align}
	\mathcal{L} \supset \frac{1}{2} \bar{f}(C_V^f \gamma^\mu + C_A^f \gamma^\mu \gamma^5)f Z_\mu + \frac{1}{2} \bar{f} (C_V^{'f} \gamma^\mu + C_A^{'f} \gamma^\mu \gamma^5) f A'_\mu\,.
$$%\end{align}
The vertex factors of the models, listed in Table~\ref{ChargeTable}, namely Model C, D, E, F, G and $B-L$ (including the minimal $B-L$ as well), can be read off from the entries in Table~\ref{tab:vertex} by plugging the values of the corresponding charges $Q'_u$ and  $Q'_d$, given in Table~\ref{ChargeTable}. The vertex factors listed in Table~\ref{tab:vertex} can also be reduced to the SM ones by taking the limiting values; $Q'_{u,d}\to 0$ and  $\epsilon\to 0\ (\xi\to 0)$. In this limit, one can see that any dark photon vertex vanishes as expected.

\section{Coherent Elastic Neutrino Nucleus Scattering (CE$\nu$NS)}
\label{nes}
	 Coherent elastic neutrino-nucleus scattering (CE$\nu$NS) includes low-energy neutrino coupling and provides an accessible window to many areas of
	physics research especially for physics beyond the Standard Model.
	CE$\nu$NS is suitable for analyzing the nonminimal dark sector framework. When scattering of neutrinos from nuclei at rest is considered, if incident neutrino energy is below 50 MeV, the interaction occurs coherently, meaning that the neutrinos interact with the nuclei as a whole rather than with its constituents individually \cite{Freedman:1973yd}. In the SM, the cross section for elastic scattering is two orders of magnitude larger than inelastic scattering, which makes CE$\nu$NS viable for observation. If a new neutral current interaction mediated by a light vector boson exists, it would not be suppressed by SM interactions in this region.  In spite of the fact that the earliest experimental proposal to measure \cevens was rather old \cite{Drukier:1984vhf}; it took lmost four decades to be able to make significant progress on the way of measuring \cevens cross section, eventually succeeded by the COHERENT Collaboration in 2017 \cite{COHERENT:2017ipa}. Hence CE$\nu$NS has become one of the important probes for physics beyond the SM since then.

The differential cross section for CE$\nu$NS is well established in the literature. The SM predicts a coherent elastic scattering cross section proportional to the weak nuclear charge, $Q_W^2$. For spin-0 and spin-1/2 targets, the differential CE$\nu$NS cross section with respect to the nuclear recoil energy $T$, in the coherent limit where the form factor approaches  unity, is given by \cite{Lindner:2016wff}
\begin{align}
\frac{d\sigma_{\textrm{SM}}}{dT}=& \frac{G_F^2 Q_W^2 M}{4\pi}\left( 1 -\frac{T}{E_\nu}-\frac{MT}{2E_\nu^2}+2 J_N \frac{T^2}{E_\nu^2}\right),\label{cross2}
\end{align}
where $G_F$ is the Fermi coupling constant, $M$ is the nucleus mass, $J_N=0,1/2$ is the spin of the nucleus, $E_\nu$ is the incident neutrino energy and $Q_W$ is the weak nuclear charge \cite{Scholberg:2005qs,Anderson:2011bi} given by
\begin{equation}
	Q_W = (2Z+N)g_V^u +(2N+Z)g_V^d = N - (1-4s_W^2)Z \label{weakcharge}
\end{equation}
where $g_V^u$ and $g_V^d$ are vector couplings of u and d quarks, $Z$ and $N$ are the atomic and neutron number respectively, and $s_W$ is the sine of the weak mixing angle. The differential cross section is expressed in nuclear recoil energy because detectors measure this quantity. The effect of the extra $T^2/E_\nu^2$ term appearing in the spin-$\frac12$ case is negligible. Therefore it is convenient to work on the spin-$\frac12$ case since more developed computational tools are available for fermionic particles. The same applies to the dark sector extended models.  

%In the $\epsilon \rightarrow 0$ and $\phi_1 \rightarrow 0$ limit of the model \eqref{LGaugeKE} - \eqref{LGauge}, the differential cross section takes the following forms:
%\begin{align}
%	\frac{d\sigma}{dT}\Big|_{\textrm{spin-}0}^{\textrm{SM+DS}}= &\frac{F^2(q^2)M}{8\pi(2MT+m_{A'}^2)^2} \left(1-\frac{T}{E_\nu}-\frac{MT}{2E_\nu^2}\right)\nonumber\\
%	&\times \left[2g_{B-L}^2(N+Z)^2 + \sqrt{2}G_F Q_W (2MT+m_{A'}^2)\right]^2,\\
%	\frac{d\sigma}{dT}\Big|_{\textrm{spin-}1/2}^{\textrm{SM+DS}}=& \frac{F^2(q^2)M}{8\pi(2MT+m_{A'}^2)^2}\left(1-\frac{T}{E_\nu}-\frac{MT}{2E_\nu^2}+\frac{T^2}{E_\nu^2}\right)\nonumber\\ 
%	&\times \left[2g_{B-L}^2(N+Z)^2 + \sqrt{2}G_F Q_W (2MT+m_{A'}^2)\right]^2,
%\end{align}

The differential cross section with respect to the nuclear recoil energy $T$ in the minimal $B-L$ model is given by
\begin{eqnarray}
\hspace*{-0.6cm} \frac{d\sigma_{BL}^{min}}{dT}&=& \frac{F^2(q^2)M}{8\pi \cos^4\!\epsilon M_Z^4 (2MT+M_{A'}^2)^2}\left(1-\frac{T}{E_\nu}-\frac{MT}{2E_\nu^2}+2 J_N\frac{T^2}{E_\nu^2}\right) \nonumber\\
	&&\times \bigg[ \left( 2MT+M_{A'}^2\right)\left(\sqrt[4]{2}\sqrt{G_F} M_Z \left(\cos\xi \cos\epsilon+ \sin\xi \sin\theta_W \sin\epsilon\right)-g_{BL} \sin\xi\right)\nonumber\\
	&&\times\left(-2A g_{BL} \sin\xi+\sqrt[4]{2}\sqrt{G_F} M_Z  \left( \cos\xi \cos\epsilon\, Q_W+ \sin\xi \sin\theta_W \sin\epsilon\, (A+2Z) \right)   \right)\nonumber\\
	&&+M_Z^2\left(\sqrt[4]{2}\sqrt{G_F} M_Z \left(\sin\xi \cos\epsilon + \cos\xi \sin\theta_W \sin\epsilon\right)+g_{BL} \cos\xi\right)\nonumber\\
	&&\times\left( 2Ag_{BL}\cos\xi +\sqrt[4]{2}\sqrt{G_F} M_Z \left(Q_W\, \cos\epsilon \sin\xi  + (A+2Z) \cos\xi \sin\theta_W \sin\epsilon\right)  \right) \bigg]^2
\end{eqnarray}
where $g_D=g_{BL}$ and $A=N+Z$ is the so-called atomic mass number of the nucleus. $F(q^2)$ is the Helm-type nuclear form factor \cite {LEWIN199687} and  $q^2$ denotes the squared momentum transfer given by $q^2=2MT$. The Helm form factor is
\begin{align}
	F(q^2) = \frac{3 j_1 \br{qR_1}}{qR_1}\ e^{-qs}
\end{align}
where $j_1(x)$ is the spherical Bessel function and $q=\sqrt{q_\mu q^\mu}$ with $q^2$ defined above, $s \approx 0.9\, \textrm{fm}$ is the nuclear skin thickness and $R_1$ is the effective nuclear radius
\begin{align}
	R_1 \simeq \sqrt{R_A^2 + \frac{7}{3}\pi^2 r_0^2 - 5s^2}
\end{align}
where
\begin{align}
	R_A \simeq \br{1.23 A^{1/3} - 0.6}\, \textrm{fm}, \quad r_0 \simeq 0.52\, \textrm{fm}\,.
\end{align}
The differential cross section expression for the 2HDM case is rather long and is not illuminating to present here.

\section{COHERENT Data for CE$\nu$NS}
\label{cohdata}

COHERENT is a neutrino-based fixed target experiment. Neutrino beams are produced by striking proton beams of pulse $\sim 1~ \mu\textrm{s}$ to a mercury target at 60 Hz,  creating $\approx 5 \times 10^{20}$ collisions per day. These collisions produce $\pi^-$ and $\pi^+$ as byproducts which  come to rest inside the target. Then, while negatively-charged pions are mostly absorbed by mercury, positively-charged ones decaying through the channel $\pi^+ \rightarrow \mu^+ + \nu_\mu$, which happens when the pion at rest,  results in monochromatic energy of $\approx 29.7~ \textrm{MeV}$ for $\nu_\mu$'s; they are called prompt neutrinos. The pion decay is then followed by the decay of the antimuon, $\mu^+ \rightarrow e^+ + \bar{\nu}_\mu + \nu_e$, which occurs about $\sim 2.2~ \mu\textrm{s}$ after the pion decay. Hence, $\bar{\nu}_\mu$ and $\nu_e$ are called delayed neutrinos.  Energy spectra for neutrinos stemming from this decay are continuous up to $52.8~ \textrm{MeV}$.  Each proton collision on target (POT)  produces on average $0.08$ neutrinos per flavor.

Before proceeding further, we would like to comment on the following possibility\footnote{We thank the anonymous referee for mentioning this possibility.}. So far in this study, we have assumed that the impact of new physics has been considered only on the scattering of neutrinos from the nucleus while the production channels of neutrinos remain as in the SM. However, neutrinos could also be produced through the decay of the dark photon, being created in the neutral pion decays via $\pi^0\to \gamma \gamma$ where one of the photons may be replaced by a dark photon. The critical point is to judge whether the dark photon would decay within the detector, which is around twenty meters away from where neutrinos are produced. Especially, this possibility would be more relevant for the low dark photon mass region ($\sim$ keV). 
	
	This requires to estimate the mean decay length of the dark photon in the laboratory reference frame. For the low mass region, if 
	the decay width of $A^\prime (\to \nu_\ell \bar{\nu}_\ell)$ is calculated (no other channel is open), the mean decay length, in the minimal $B-L$ scenario, is around $10^3$ km for the parameters values  $M_{A^\prime}=1$ keV and $(g_{B-L},\sin\epsilon)=(10^{-5},10^{-4})$. A similar calculation for the other models in our study (Model C -- Model G) is carried out and found that the  mean decay length of the dark photon in the low mass region is few times longer than the one in the minimal $B-L$ model. Therefore, direct neutrino production from dark photons could only happen out of the detector and thus in the rest of the analysis such contributions are not going to be pursued further.

The produced neutrinos then collide with 14.57 kg CsI and 24 kg liquid argon targets located at $19~ \textrm{m}$ and $27.5~ \textrm{m}$ away from the mercury target. The energy-density expressions for incoming neutrino flux $f_ {\nu_\alpha}$ for prompt $\nu_\mu$ and delayed $\nu_e$ and $\bar{\nu}_\mu$  beams are given by \cite {Amanik:2009zz, Moreno:2015bta}:
\begin{align}
	f^{\nu_\mu}_E(E_{\nu_\mu})\propto&\; \delta \br{E_{\nu_\mu}-\frac{m_\pi^2-m_\mu^2}{2m_\pi}}\nonumber,\\
	f^{\bar{\nu}_\mu}_E(E_{\bar{\nu}_\mu})\propto&\; \frac{64E_{\bar{\nu}_\mu}^2}{m_\mu^3}\left(\frac{3}{4}-\frac{E_{\bar{\nu}_\mu}}{m_\mu} \right),\\
	f^{\nu_e}_E(E_{\nu_e})\propto&\; \frac{192E_{\nu_e}^2}{m_\mu^3}\left(\frac{1}{2}-\frac{E_{\nu_e}}{m_\mu} \right).\nonumber
\end{align}
Time-dependent flux density $f_t^{\nu_\alpha}(t)$ is presented in \cite{COHERENT:2021xhx}   and after normalizing $f_E^{\nu_\alpha}$ and $f_t^{\nu_\alpha}$, the total neutrino flux can be expressed as
\begin{align}
f^{\nu_\alpha}\br{E_\nu,t} = \mathcal{N} f_E^{\nu_\alpha}\br{E_\nu} f_t^{\nu_\alpha}(t) 
\end{align}
where $\mathcal{N}=rN_{POT}/4\pi L^2$. Here $L$ is the distance between the detector and the neutrino source, $r$ is the number of neutrinos per flavor created in each POT collision, and $N_{POT}$ is the total number of POT collisions throughout the entire experiment, which is $1.76\times 10^{23}$ for the CsI 2017 data, $3.198 \times 10^{23}$ for the CsI 2022 data and $1.37 \times 10^{23}$ for the LAr-based experiment.

Nuclear recoil energy is picked up by scintillation detectors and converted into photoelectrons (PE). This conversion is parametrized with light yield $\mathcal{L}_Y$, which is the amount of PE produced per unit energy. Values provided with the data releases are $\mathcal{L}_Y = 13.348~$$\rm PE/keV\!ee$ for CsI and $\sim 4.5~$$\rm PE/keV\!ee$ for LAr-based experiment. The number of produced PE can be expressed as
\begin{align}
	n_{\textrm{PE}} = {\rm QF} \;T\; \mathcal{L}_Y . 
\end{align}
Nuclear recoil energy is primarily dissipated through secondary nuclear recoils, and only a small amount of energy is turned into scintillation (or ionization). The Quenching factor QF is the ratio of energy turned into scintillation
\begin{align}
{\rm QF} = \frac{E_{\textrm{ee}}}{T},
\end{align} 
where the subscript ``$\textrm{ee}$'' stands for electron equivalent and $E_{\rm ee}$ is the equivalent energy of a recoiling electron in which the energy is dissipated through only scintillation. There have been different approaches for quenching factor in each  data release by the COHERENT Collaboration. In the 2017 release,  a constant quenching factor ${\rm QF}_{\textrm{CsI}} = 8.78 \pm 1.66\%$ was suggested. Later, an energy-dependent model for the quenching factor was proposed in \cite{Collar:2019ihs}, which increases the accuracy of the SM expectation for the event rate. The energy -dependent model describes the scintillation in CsI by slow ions with low-energy approximation to Birks' scintillation model \cite{Birks:1951boa} multiplied by an adiabatic factor to account for the behavior of scintillation production cutoff at low nuclear recoil energy limit. The quenching factor takes the form 
\begin{align}
{\rm QF} (T)= \big[kB \cdot \br{dE/dr}_i\big]^{-1} \br{1-\exp \br{-T/ E_0}},
\end{align}
where $kB = 3.311 \pm 0.075 \times 10^{-3} {\rm g\, MeV^{-1}cm^{-2}}$ and $E_0 = 12.97 \pm 0.61\,  {\rm keV}$. And $\br{dE/dr}_i$ is the total stopping power for ions, extracted from the software SRIM-2013 \cite{ZIEGLER20041027}. Both {\rm QF} proposals for CsI will be considered in the analysis in the following section. 

In the recent study by the COHERENT Collaboration \cite{COHERENT:2021xmm}, the previously proposed quenching factor for CsI \cite{COHERENT:2017ipa} was reassessed by including  a new scintillation response to nuclear recoil measurement on CsI[Na] crystal. Quenching in the region of interest is modeled as a fourth-degree polynomial fit to the  available data \cite{COHERENT:2021pcd}
\begin{align}
E_{\rm ee} = g(T) = 0.05546\,T + 4.307\,T^2 - 111.7\,T^3 + 840.4\,T^4
\end{align}
where the detector response $E_{\rm ee}$ is in MeVee and the nuclear recoil energy $T$ is in MeVnr.

For the LAr-based experiment, the suggested quenching factor is a linear fit to the world data on argon which is given in Fig.~7 of \cite{COHERENT:2020iec}.

Collision events are counted in $n_{\textrm{PE}}$ and time bins. 
For the CsI 2017 and LAr releases, the expected number of events in the $i$th PE and $j$th time bin can be written as
\begin{align}
N_{event}^{i,j}=\sum_{\alpha = {\rm flavor}}\sum_{\beta={\rm Nucleus}} N_{targ}^\beta \int^{T^i_{\rm max}}_{T^i_{{\rm min}}}  \int^{E_\nu^{\rm max}}_{E_\nu^{\rm min}} \int^{t_{\rm max}^j}_{t_{\rm min}^j} f^{\nu_\alpha}(E_\nu, t)\mathcal{A}\br{T}\, \frac{d\sigma}{dT}\,dt\, dE_\nu\,  dT \label{theornumevnt}
\end{align}
where $N_{targ}^\beta$ is the number of nuclei in the target, $E_\nu^{\rm min}$ and $E_\nu^{\rm max}$ are energy limits of the incident neutrinos with $E_\nu^{\rm max} \approx 52.8\, {\rm MeV}$, and $T_{\rm min}^i$ and $T_{\rm max}^i$ give the boundary values in energy for the $i$th bin. Lastly $\mathcal{A}$ is the signal acceptance function \cite{COHERENT:2018imc},
\begin{eqnarray}
\mathcal{A}(n_{\rm PE})=  \frac{a}{1+\exp[-k(n_{\rm PE}-x_0)]}\Theta(n_{PE})
\end{eqnarray}
where the parameters have the following values
\begin{eqnarray}
a=0.6655^{+0.0212}_{-0.0384},\nonumber\\
k=0.4942^{+0.0335}_{-0.0131},\nonumber\\
x_0=10.8507^{+0.1838}_{-0.3995}
\end{eqnarray}
and $\Theta(n_{PE})$ is a modified Heaviside step function defined as 
\begin{eqnarray}
\Theta (n_{PE})=
\begin{cases}
0 & n_{PE}<5\\
0.5 & 5 \leq n_{PE} < 6\\
1 & n_{PE}\geq 6\,.
\end{cases}
\end{eqnarray}

In the CsI 2022 analysis of the COHERENT Collaboration, functions for energy smearing and acceptance for both energy and time spectra are provided  \cite{COHERENT:2021xmm}. 
	Acceptance in the energy spectrum is
	\begin{align}
		\mathcal{A}_\pe\br{\npe}=\frac{1.32045}{1+ \exp\!\left[-0.285979 (\npe - 10.8646)\right]} -0.333322
	\end{align}
and the time-dependent part of the acceptance function is given by
\begin{align}
\mathcal{A}_t\br{t} =  \begin{cases}
1 & t< a\\
e^{-b \br{t-a}} & t \geq a
\end{cases}
\end{align}
where 
\begin{align}
a &= 0.52 \mu s\nonumber\\
b &= 0.0494 /\mu s,
\end{align}
and the energy  smearing is parametrized in the following form by using the gamma function $\Gamma(1+b)$
\begin{align}
P\br{\npe^{\rm reco} \big| E_{\rm ee}^{\rm true}} =  \frac{\left[a (1+b)\right]^{1+b}}{\Gamma(1+b)}(\npe)^b e^{-a(1+b)\npe}
\end{align}
where ``true" stands for the true spectrum which would be the one  obtained without the smearing effect, and ``reco" stands for the measured spectrum. Here the parameters $a = 0.0749/E_{\rm ee}^{\rm true}$ and $b = 9.56 \times E_{\rm ee}^{\rm true}$ depend on the quenched-energy deposition. The energy smearing is normalized by using the following condition 
\begin{align}
\int_{\Omega^{\rm reco}} dT^{\rm reco} P\br{\npe^{\rm reco}\br{T^{\rm reco}}|E_{\rm ee}^{\rm true}} = 1\,.
\end{align}
With these effects taken into account, the number of events in the $i$th PE and $j$th time bin is given by 
\begin{eqnarray}
N_{\rm event}^{i,j}&=&\sum_{\beta={\rm Nucleus}}N_{targ}^\beta \sum_{\alpha = {\rm flavor}} \int^{T^i_{\rm max}}_{T^i_{{\rm min}}} \int^{E_{\rm ee}^{\rm max}}_{E_{\rm ee}^{\rm min}} \int^{E_\nu^{max}}_{E_\nu^{min}} \int^{t^j_{\rm max}}_{t^j_{\rm min}} dT^{\rm reco}  dE_{\rm ee}^{\rm true} dt dE_\nu   \nonumber\\
&&\times f^{\nu_\alpha}(E_\nu, t) \mathcal{A}_{\rm PE}(T^{\rm reco})  \mathcal{A}_{t}(t) P\br{n_{\rm PE}\br{T^{\rm reco}}\big| E_{\rm ee}^{\rm true}}
 \frac{d\sigma}{dT}\br{T^{\rm true}\Big|_{T= g^{-1}\br{E_{\rm ee}} }}.
\end{eqnarray}

\begin{table*}[htb]
	\caption{Our calculated values for the total number of events in the SM, in comparison to the literature}
	\label{table:Nevents}
	\begin{tabularx}{0.99\textwidth}{@{\extracolsep{\fill{}}}lccc}
		\hline\hline
		\qquad\qquad Data Set & Our Study & Literature & References\\
		\hline
		CsI 2017 (constant QF) & 173 & 173 & \cite{COHERENT:2017ipa}\\
		CsI 2017 (energy-dependent QF) & 139 & 138 & \cite{Collar:2019ihs}\\
		CsI 2022 & 437 & 431 & \cite{COHERENT:2021xmm}\\
		LAr$-$Analysis A & 128 & 128 & \cite{COHERENT:2020iec}\\
		LAr$-$Analysis B & 101 & 101 & \cite{COHERENT:2020iec}\\
		\hline\hline
	\end{tabularx}
\end{table*}

At the end of the section, 	our findings for the total number of events in  the SM  are listed in Table \ref{table:Nevents} and some of the results available in the literature are also added for comparison.

\section{Statistical Analysis}
\label{analysis}
 In this section, we will adopt a $\chi^2$-fit to study the sensitivity of the COHERENT data  to phenomenological parameters in the framework of new physics interactions. For the current analysis we define the following  $\chi ^2$ function depending on a parameter set $\cal P$
 \begin{align}
	\chi ^2 ({\cal P}) = \sum_{i = {\rm bins}} \frac{\left(N_{\rm meas}^i-N_{\rm exp}^i(1+\alpha)-B_{\rm on}^i[1+\beta]\right)^2}{\left(\sigma^{i}_{\rm stat}\right)^2}+ f_{\rm pull}\br{\alpha, \sigma_\alpha} + f_{\rm pull}\br{\beta, \sigma_\beta}
	\label{chi2mb}
\end{align}
where $N_{\rm meas}^i$ and $N_{\rm exp}^i$ are measured and expected number of events in the $i$th bin respectively, $B_{\rm on}^i$ is the estimated number of background events when the beam is on, $\alpha$ and $\beta$ are the systematic parameters for the signal rate and $B_{\rm on}$, respectively. The $\chi^2$ function is minimized over $\alpha$ and $\beta$. Here, the statistical uncertainty of the measurement is given by $\sigma_{\rm stat}^i=\sqrt{N_{\rm meas}^i+B_{\rm on}^i+2B_{\rm SS}^i}$, where  $B_{\rm SS}^i$
denotes the steady-state backgrounds. The so-called pull terms presented by the COHERENT Collaboration \cite{COHERENT:2017ipa} have the form
\begin{align}
f_{\rm pull}\br{x, \sigma_x} = \left( \frac{x}{\sigma_x}\right)^2\,,\quad x=\alpha,\beta
\end{align}
where $\sigma_\alpha$ and $\sigma_\beta$ are the fractional uncertainty of $\alpha$ and $\beta$, corresponding to 1-sigma variation. The pull term in the above form is observed to lead to unphysical results around its limiting values. This behavior is also noted in \cite{Denton:2020hop} and instead, an asymmetric pull term of the form
\begin{align}
f_{\rm pull}\br{x, \sigma_x} = \frac{2}{\sigma_x^2}\br{x -\log \br{x+1}},
\end{align}
is suggested to use. For completeness, we use both forms of the pull terms in the $\chi^2$ calculation and comment on their effect.

Even though the scattering data are obtained in a multibinned detector, the earlier analyses provided by the COHERENT Collaboration combined all events in a single bin \cite{COHERENT:2017ipa} for CsI 2017 and LAr data. Later it is suggested to adopt rather a multibin analysis and indeed in the latest study by the COHERENT Collaboration more than one multibin options have been performed. The $\chi ^2$ function for the single bin analysis can be obtained from Eqn.~(\ref{chi2mb}) by simply using the total number of events for the signal and the background.

\subsection{Numerical Discussion}

In this study it is promised to analyze various two-Higgs-doublet models extended by a dark $U(1)_D$ gauge symmetry with the use of the \cevens data measured by the COHERENT Collaboration. This will be achieved by looking at the effects of various parameters and factors present in the calculation.   $M_{A'}$,  $\sin\epsilon$,  $g_D$, $\tan\beta$, $Q_u^\prime$, $Q_d^\prime$ are the parameters of the theoretical framework considered here. There are additional factors like the number of bins used (1PE - 1t time bin, 9PE-1t and 9PE-10t bins), data taken at different times for various targets, which we call CsI 2017, CsI 2022, LAr A and LAr B, and  several approaches for the quenching factor. There are totally seven chosen representative scenarios, obtained by fixing $(Q_u^\prime$, $Q_d^\prime)$ values. Taking into account the number of parameters, factors and the models mentioned above, it would not be  feasible to present all possible plots here. Instead, we will plot two distributions for each model in the $(g_D,\sin\epsilon)$ and $(g_D,M_{A^\prime})$ planes by varying one factor at a time while keeping the rest at their best settings which are proven to provide the most stringent constraint on the parameter space.
Models showing very similar behaviors  and cases with parameters or factors displaying not much sensitivity in the $(g_D,\sin\epsilon)$ or  $(g_D,M_{A^\prime})$ plane  are not presented here.

%********************MINIMAL MODEL********************

%		\fgruler{upperleft}{2.75cm}{2cm}
\begin{figure*}[htbp]
	$\begin{array}{cc}
		\hspace*{-0.3cm}
		\includegraphics[scale = 0.104]{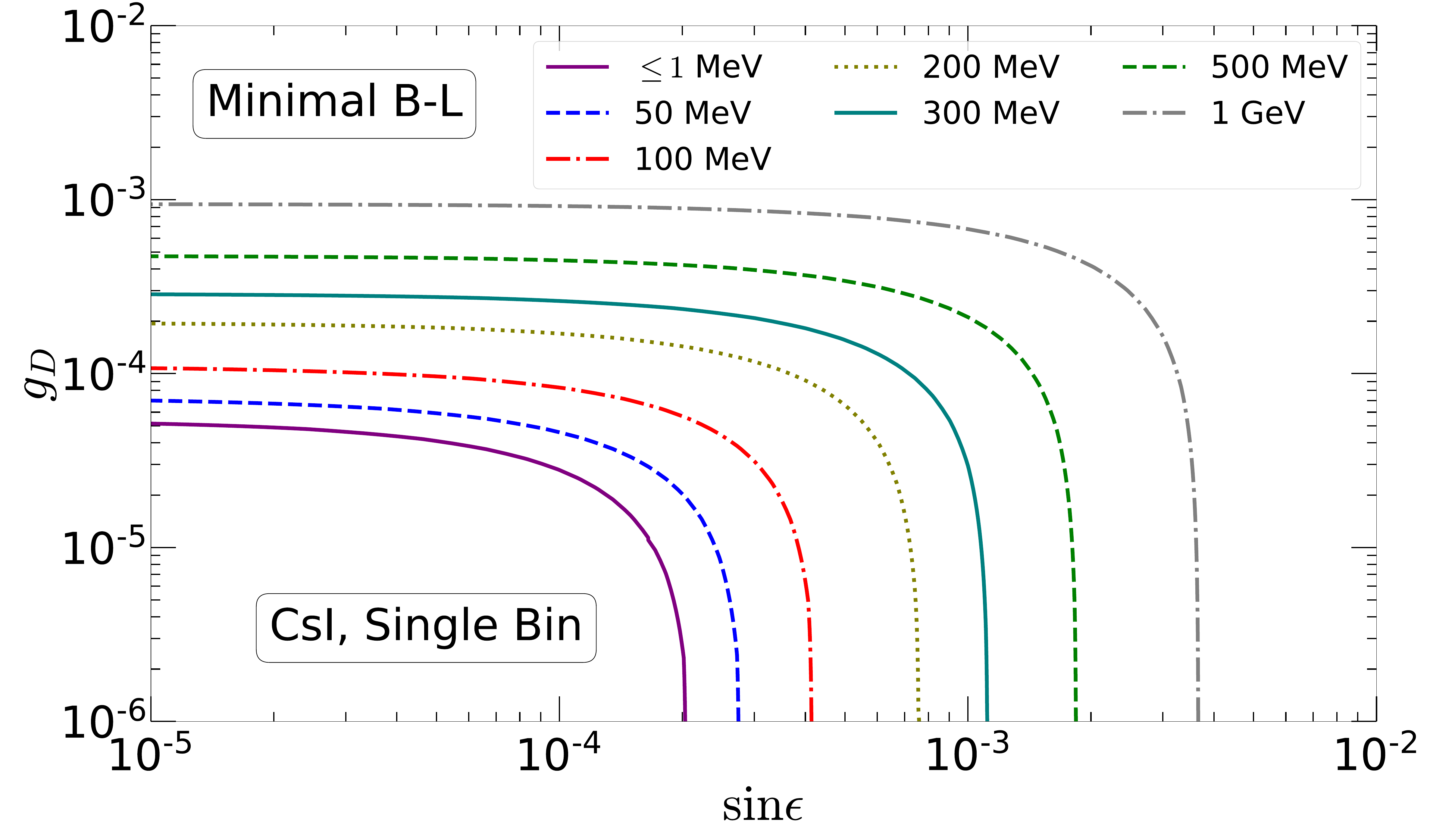} & 
		\includegraphics[scale = 0.104]{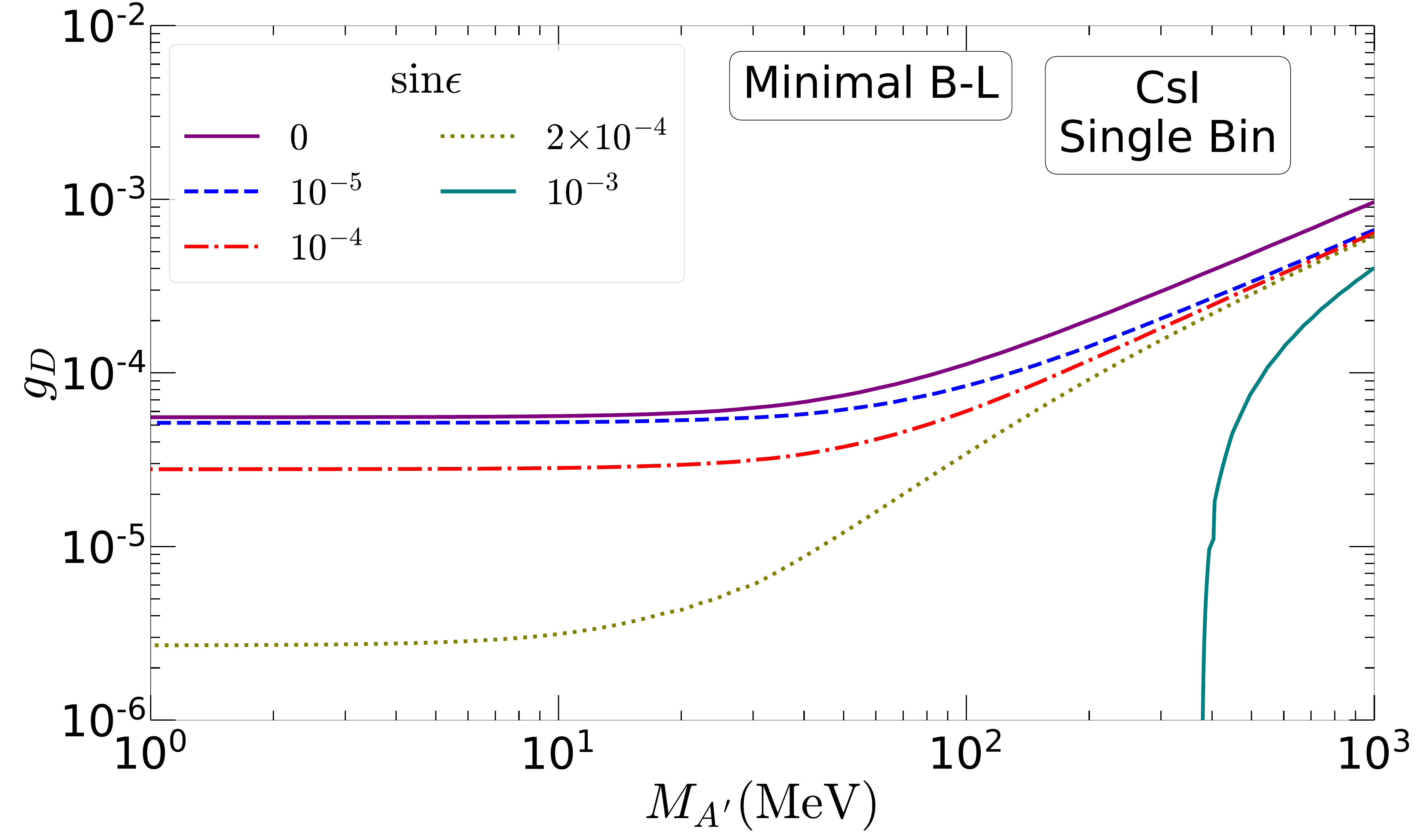}
	\end{array}$
	\vskip -0.4cm
	\caption{The exclusion curves in the $(g_D,\sin\epsilon)$ and $(g_D,M_{A'})$ parameter spaces for various $M_{A'}$ (left) and $\sin\epsilon$ (right) values, respectively, in the minimal $B-L$ model. Regions above the curves are excluded with $90\%$ C.L. by the COHERENT data for \cevens\!\!.}
	\label{fig:minBL}
\end{figure*}

In the left panel of Fig.~\ref{fig:minBL}, the exclusion curves in the $(g_D,\sin\epsilon)$ parameter space for various $M_{A'}$ values   are shown for the minimal $B-L$ model.  $M_{A'}$ values are chosen to cover light-dark photon masses as well as masses up to $1$ GeV. Regions above the curves are excluded with $90\%$ C.L. by the COHERENT data for \cevens\!\!. It is seen that the best bound on the parameters  $g_D$ and $\sin\epsilon$ is obtained for the light dark photon scenario ($M_{A'}\leq 1$ MeV). This is expected since lighter dark photon mass makes the new physics contributions to the cross section of the \cevens bigger, exceeding the observed number of events. Hence, this would force the parameters $g_D$ and $\sin\epsilon$ to be smaller. 

In the right panel of Fig.~\ref{fig:minBL}, similar exclusion curves are given in the $(g_D,M_{A'})$ parameter space for different $\sin\epsilon$ values, including no kinetic mixing scenario.  To be consistent with the observed number of events at $90\%$ CL, larger kinetic mixing pushes the bound on $g_D$ to smaller values. The sensitivity to the kinetic mixing parameter starts to show up at around $M_{A'}\sim 100$ MeV. 

\begin{figure*}[htbp]
	$\begin{array}{cc}
		\hspace*{-0.3cm}
		\includegraphics[scale = 0.104]{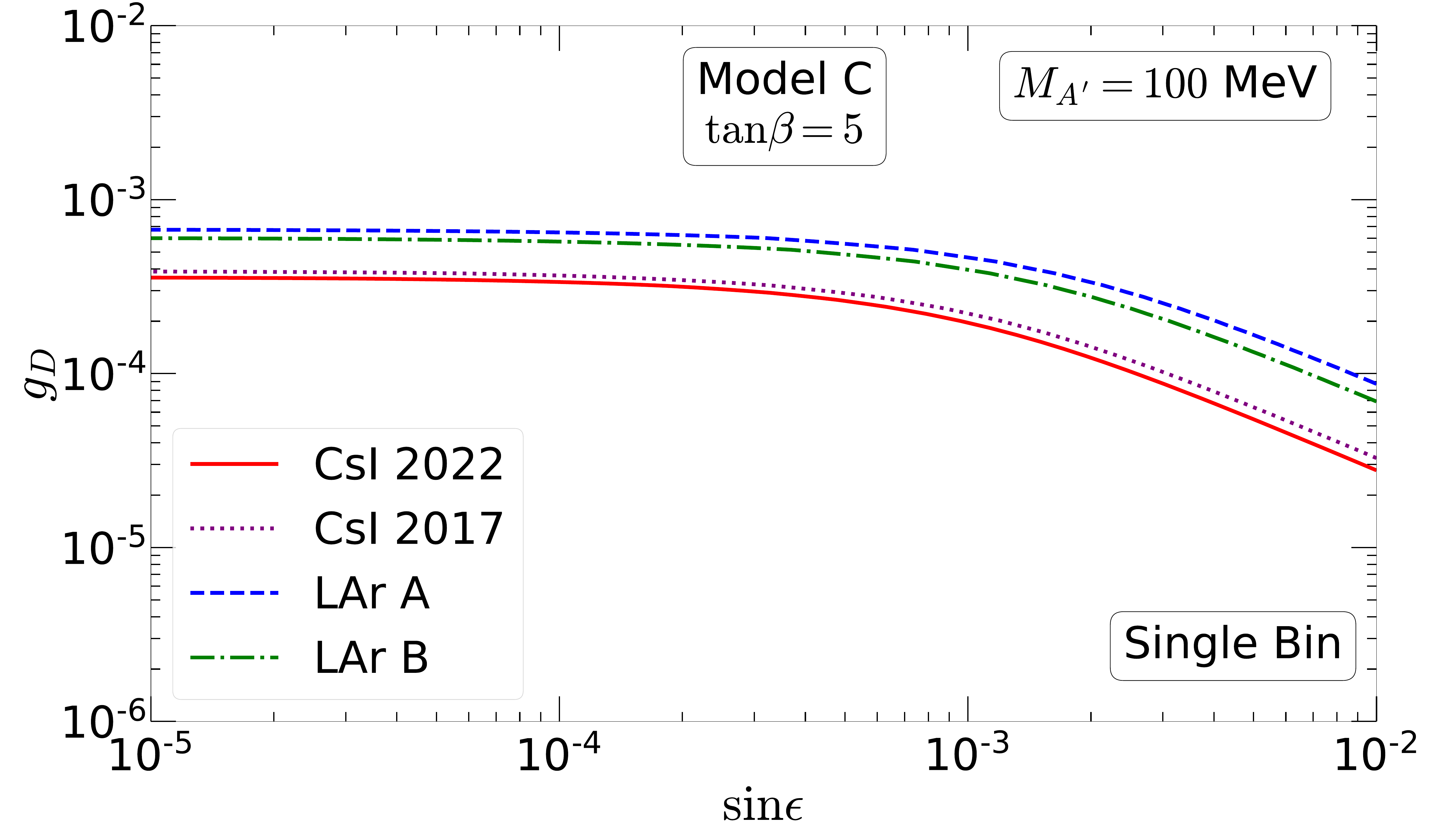} &
		\includegraphics[scale = 0.104]{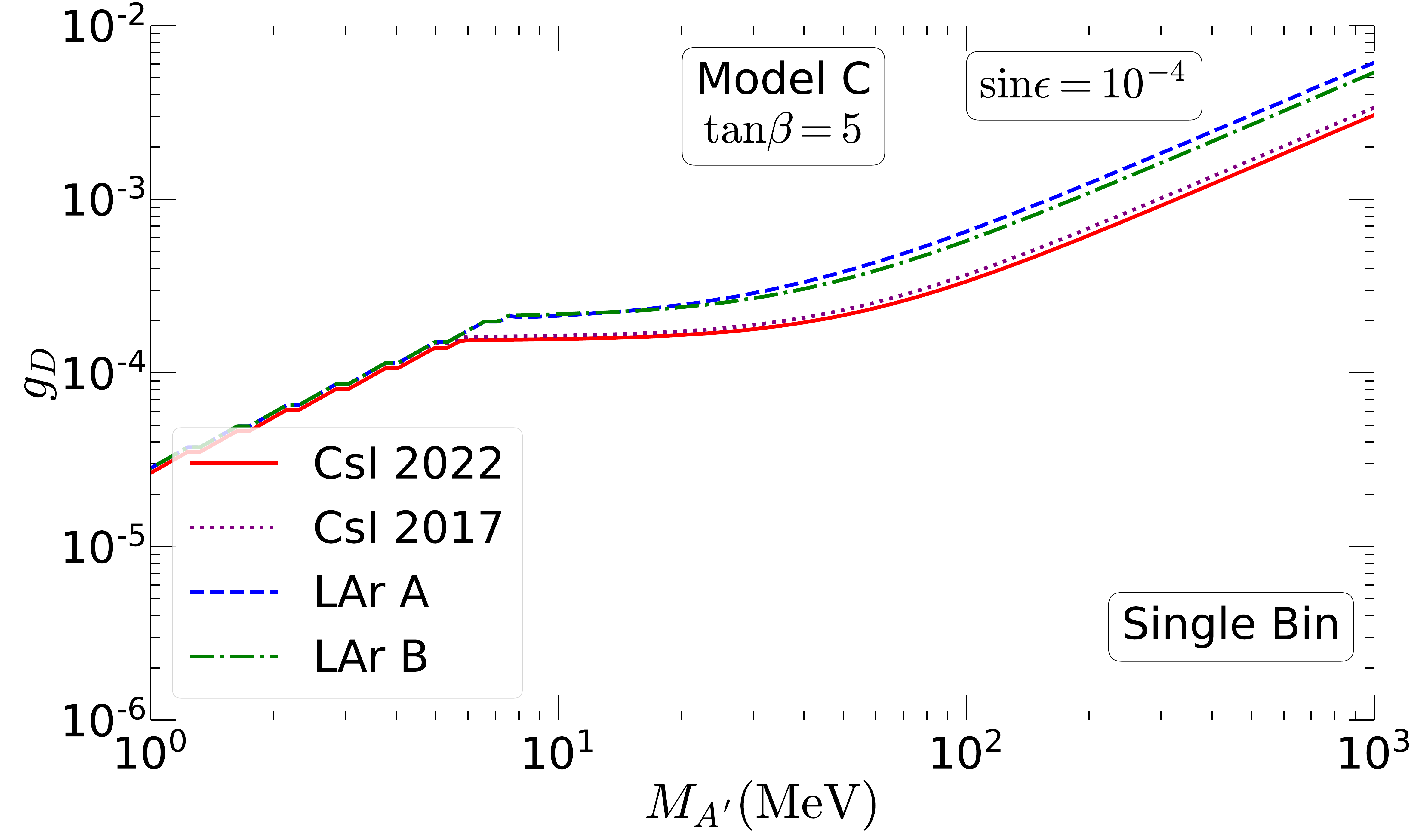}
	\end{array}$
	\vskip -0.4cm
	\caption{The exclusion curves in the $(g_D,\sin\epsilon)$ parameter space for $M_{A'}=100$ MeV (left) and $(g_D,M_{A'})$ parameter space for $\sin\epsilon=10^{-4}$ (right) for the sources CsI 2017, CSI 2022, LAr option A and LAr option B in the Model C. Regions above the curves are excluded with $90\%$ C.L. by the COHERENT data for \cevens\!\!.}
	\label{fig:modC}
	\end{figure*}

\begin{figure*}[htbp]
$\begin{array}{cc}
	\hspace*{-0.3cm}
	\includegraphics[scale = 0.104]{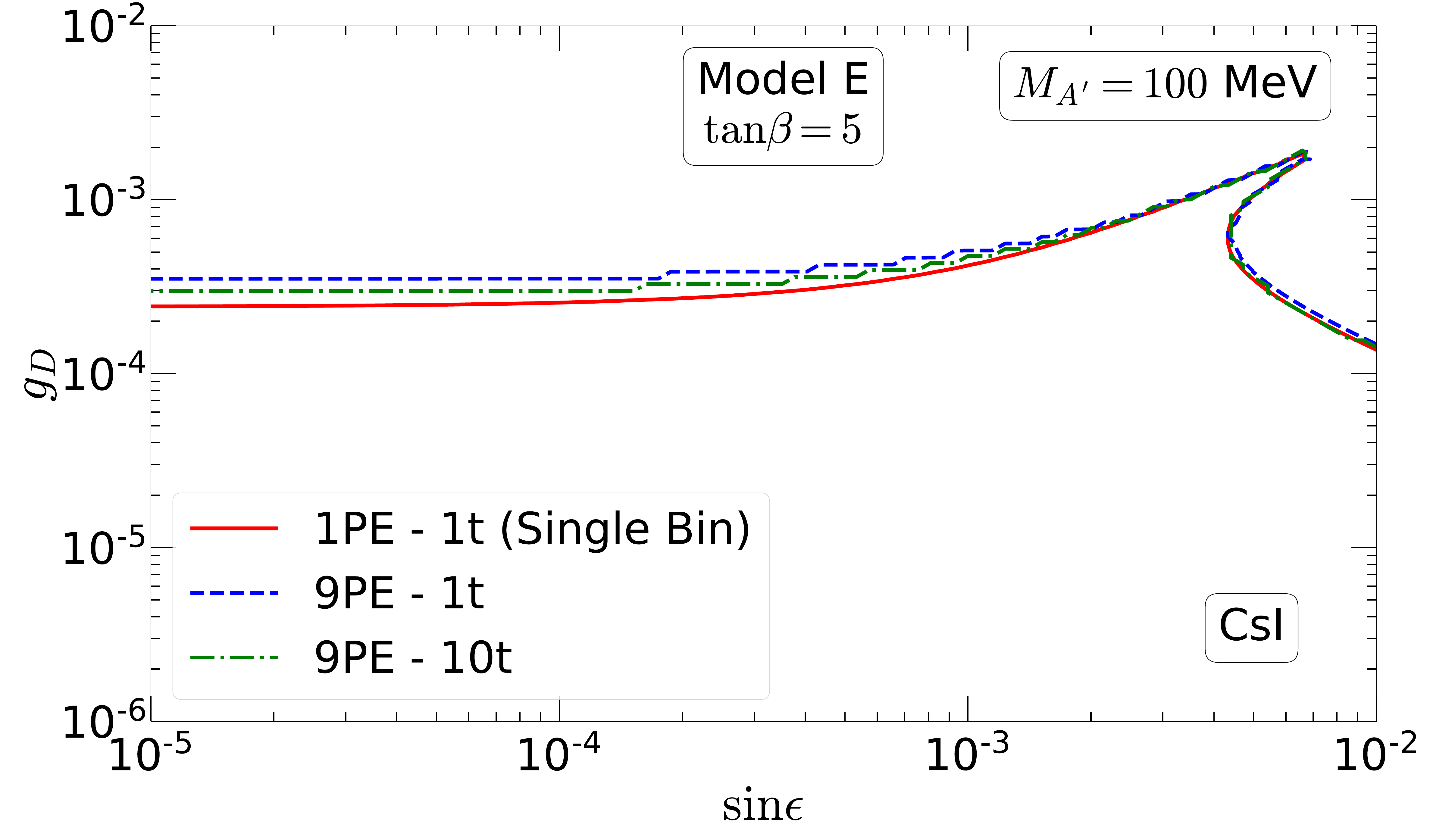} &
	\includegraphics[scale = 0.104]{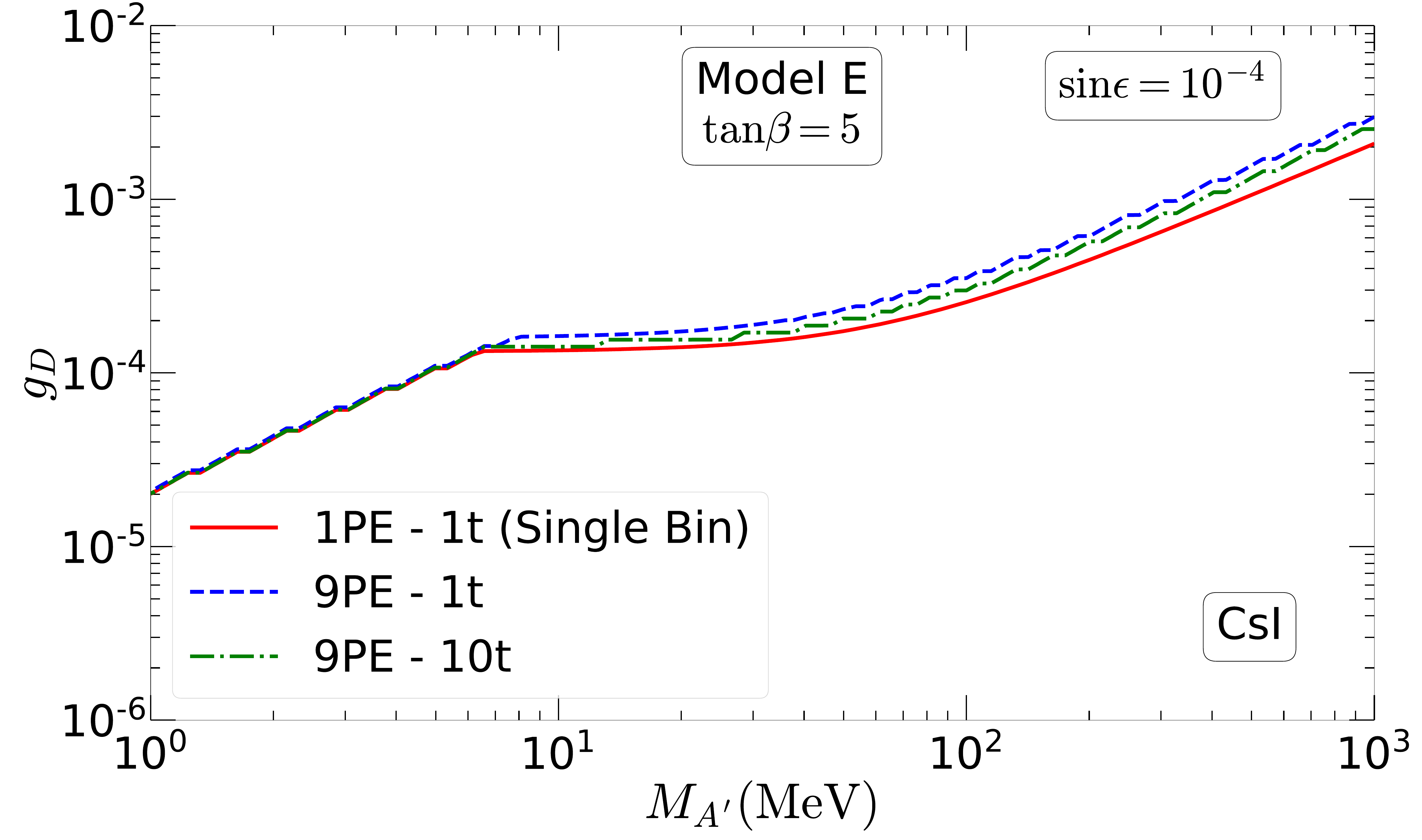}
\end{array}$
\vskip -0.4cm
\caption{The exclusion curves in the $(g_D,\sin\epsilon)$ parameter space for $M_{A'}=100$ MeV (left) and $(g_D,M_{A'})$ parameter space for $\sin\epsilon=10^{-4}$ (right) for the single-bin (1PE-1t) and multibin (9PE-1t and 9PE-10t) analyses in the Model E. Regions above the curves are excluded with $90\%$ C.L. by the COHERENT data for \cevens\!\!.}
\label{fig:modE}
\end{figure*}

The results presented in  Fig.~\ref{fig:minBL} have been obtained under the assumption that the neutrinos collide with a CsI target, the quenching factor, $QF$, is energy dependent, and the $\chi^2$ analysis has been carried out with one single  bin (1PE-1t). In fact, our search has indicated that these are the circumstances where better bounds are possible as compared to the case where the target is LAr and  the $\chi^2$ is minimized over multibins (9PE-1t or 9PE-10t). This feature is not only true for the minimal $B-L$ scenario but also valid for the rest of the models considered. Indeed, Figs.~\ref{fig:modC} and \ref{fig:modE} verify our claim specifically  within the framework of Models C and E but, in general,  this would be the case in all models.   Let us comment on the left tails of the right panels in Figs.~\ref{fig:modC} and \ref{fig:modE}. Starting around a critical $M_{A'}$ value ($\sim$ 10 MeV), different curves shrink to a single one which becomes linear for smaller $M_{A'}$ region. The boundary of the excluded part in this end is determined by the minimum mass requirement (see Fig.~\ref{fig:comparemass} and the discussion in the text) while in the heavier dark photon region the COHERENT data take over and give more stringent bounds.

\begin{figure*}[htbp]
	$\begin{array}{cc}
		\hspace*{-0.3cm}
		\includegraphics[scale = 0.104]{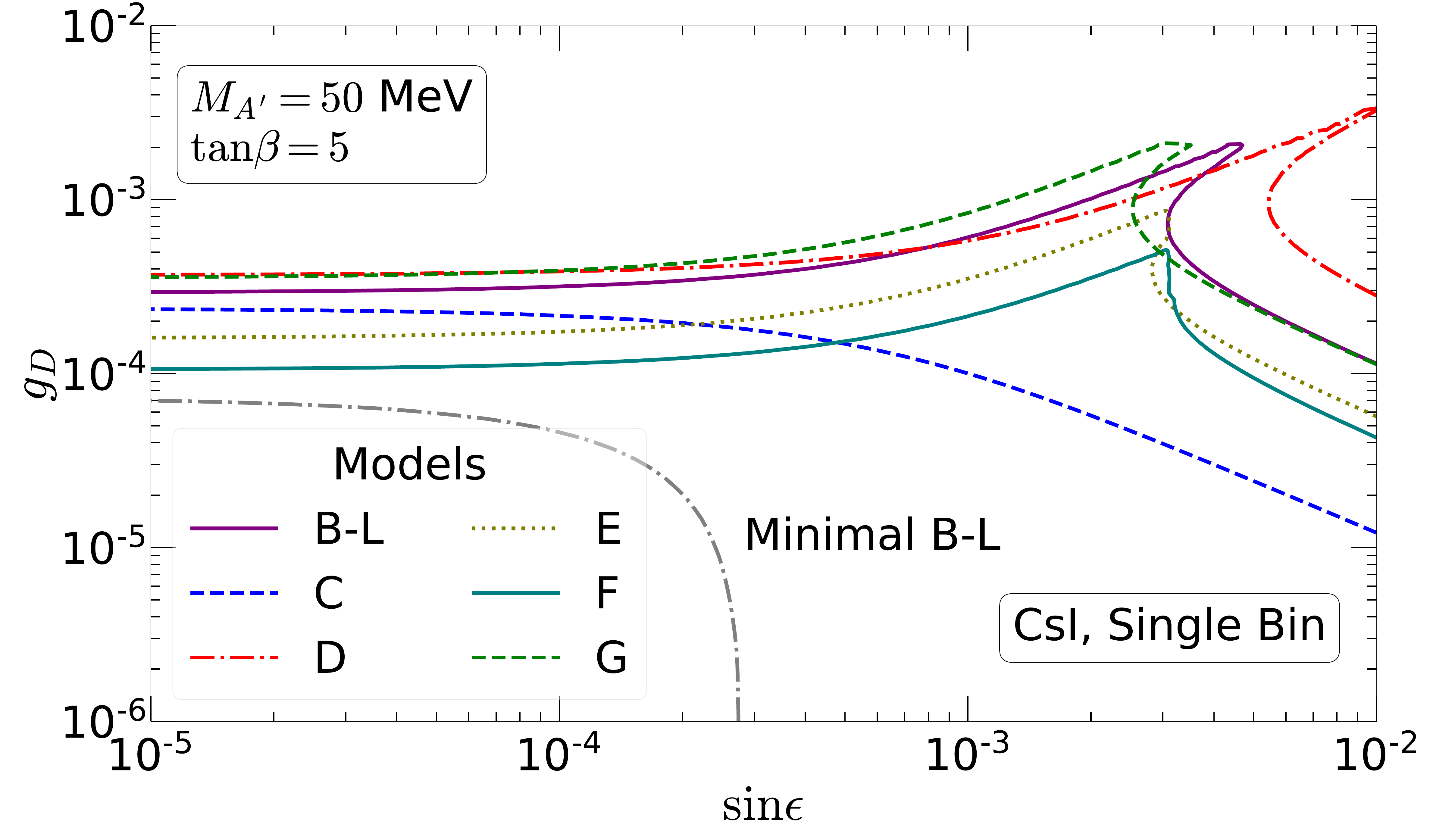} &
		\includegraphics[scale = 0.104]{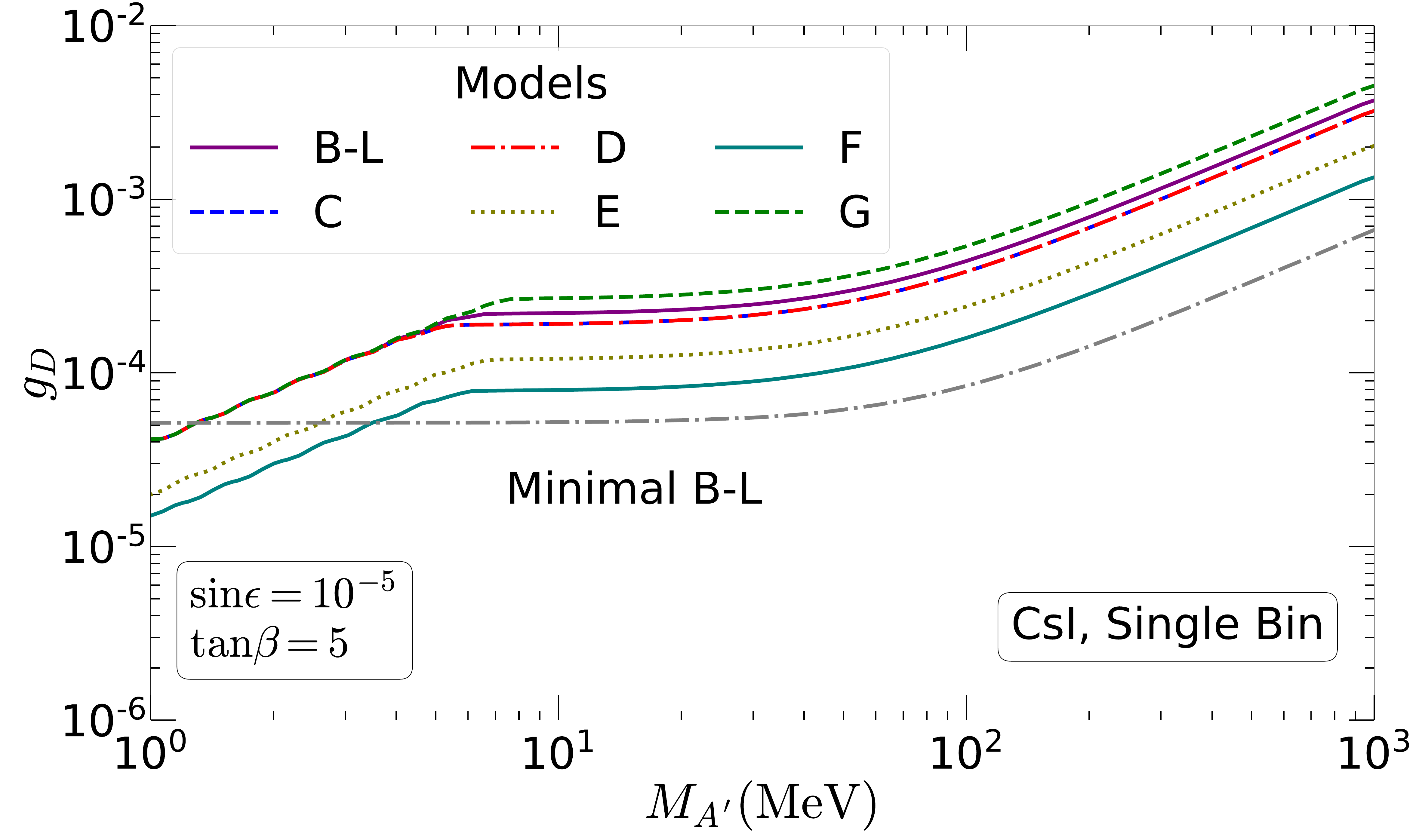}
	\end{array}$
	\vskip -0.4cm
	\caption{The exclusion curves in the $(g_D,\sin\epsilon)$ parameter space for $M_{A'}=100$ MeV (left) and $(g_D,M_{A'})$ parameter space for $\sin\epsilon=10^{-4}$ (right) for all the models considered. Regions above the curves are excluded with $90\%$ C.L. by the COHERENT data for \cevens\!\!.}
	\label{fig:modcomparison}
\end{figure*}

   \begin{figure*}[htbp]
	$\begin{array}{cc}
		\hspace*{-0.3cm}
		\includegraphics[scale = 0.107]{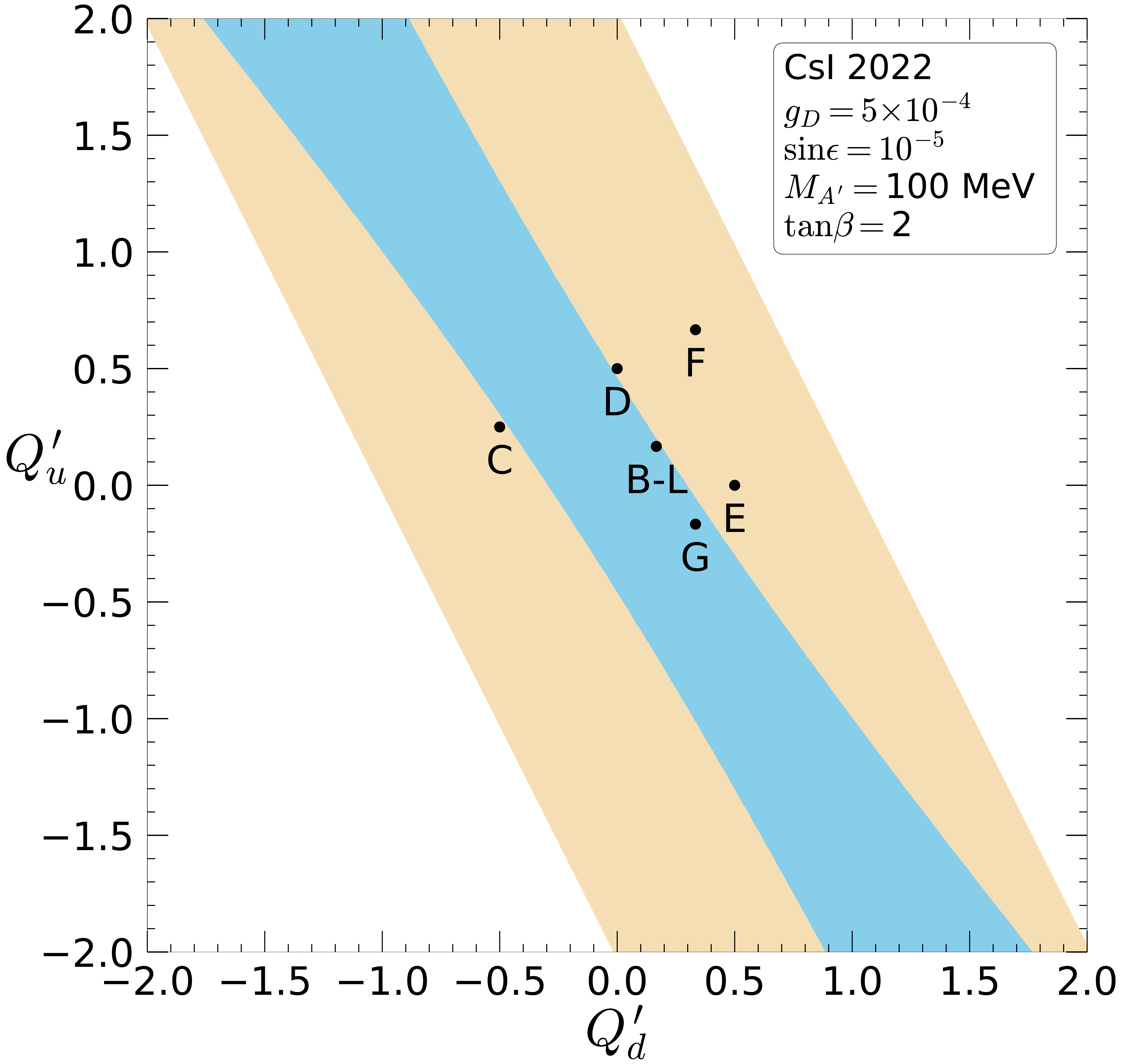} &\hspace*{-0.0cm}
		\includegraphics[scale = 0.107]{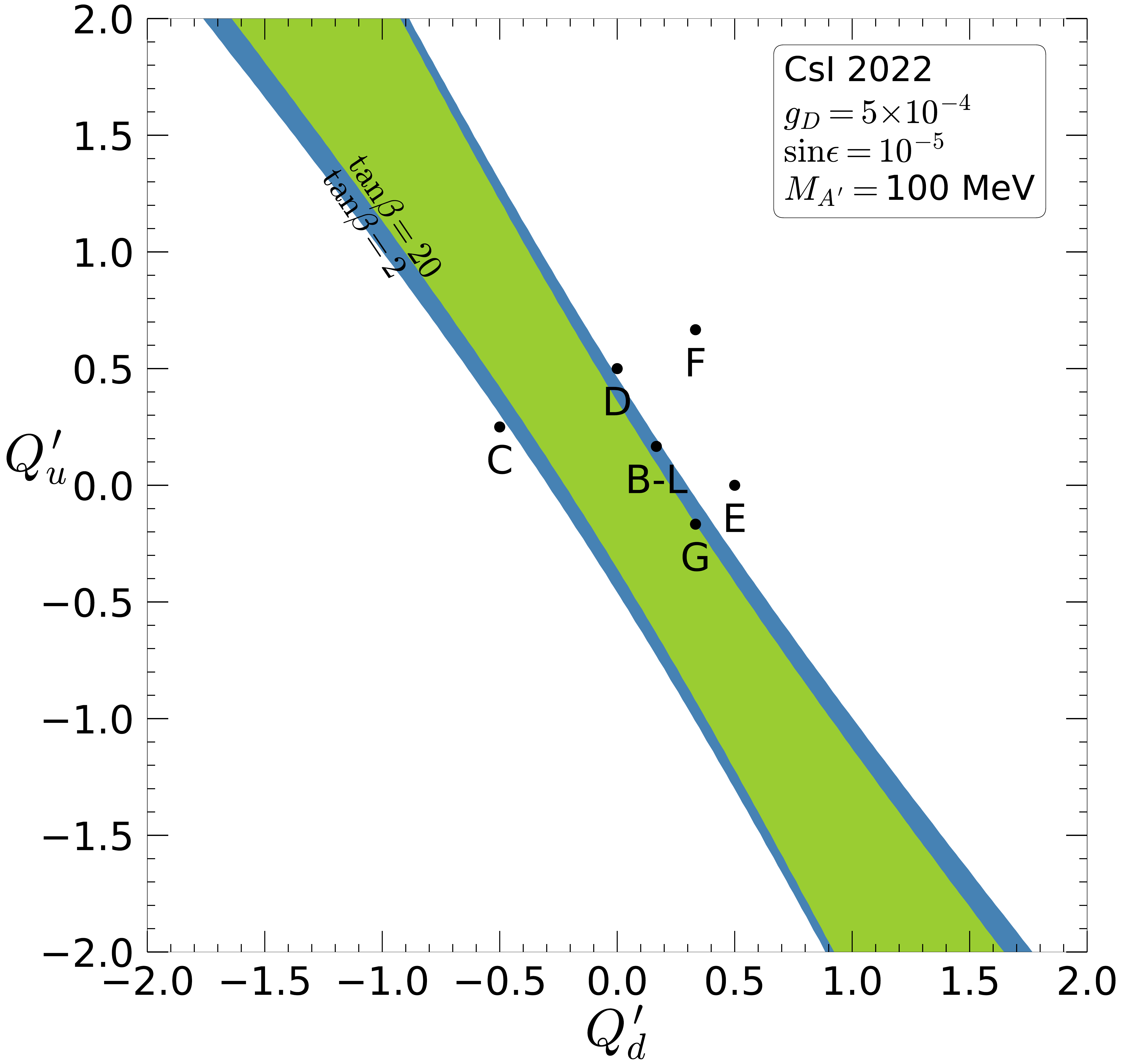} \\
		\hspace*{-0.3cm}
		\includegraphics[scale = 0.107]{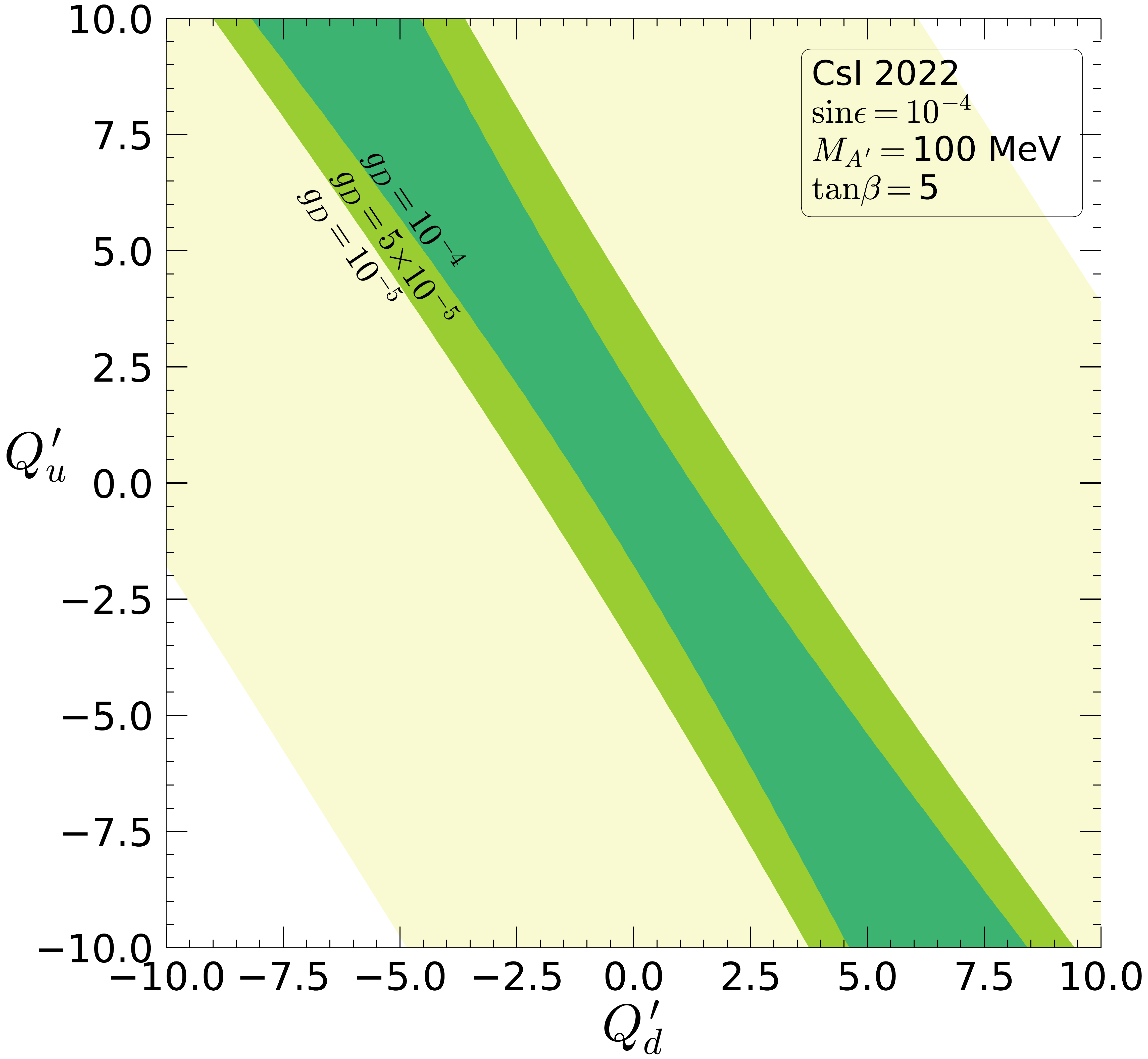} &
		\hspace*{-0.0cm}
		\includegraphics[scale = 0.107]{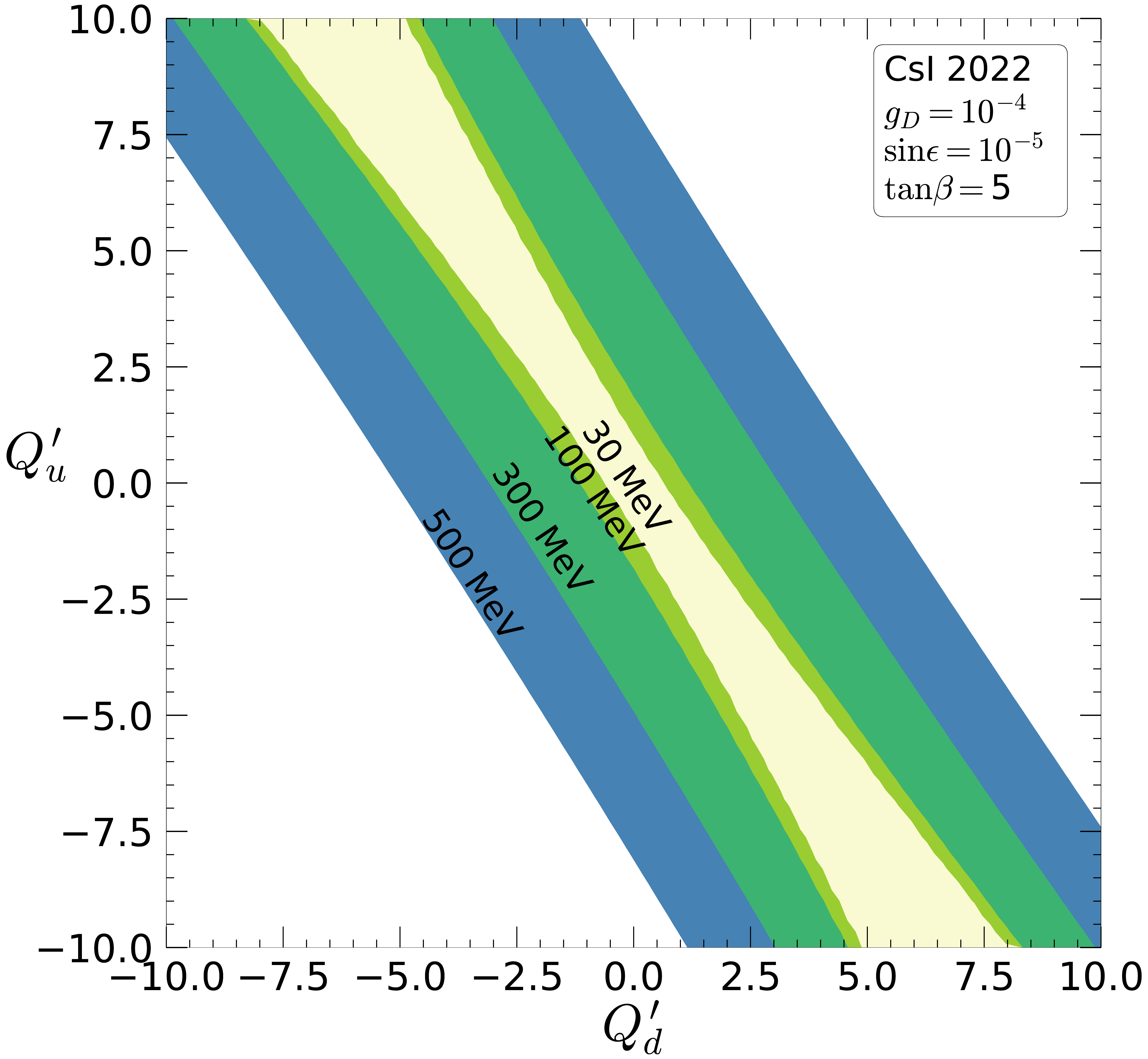} \\
	\end{array}$
	\vskip -0.4cm
	\caption{Allowed $90\%$ C.L. regions for the dark charges 	$Q_u^\prime$ and  $Q_d^\prime$ for the CsI target. In the upper row, $g_D=5\times 10^{-4},\sin\epsilon=10^{-5}$, and $M_{A'}=100$ MeV are chosen with $\tan\beta=2$ (upper left) and for two different $\tan\beta$ values (upper right). The $(Q_u^\prime,Q_d^\prime)$ values of all of the 2HDMs extended with $U(1)_D$ are marked on the graphs given in the upper row.   In the lower row, $\tan\beta=5$ with $\sin\epsilon=10^{-4}$ and $M_{A'}=100$ MeV for various $g_D$ (lower left) and $\sin\epsilon=10^{-5}$ and $g_D=10^{-4}$ for various $M_{A'}$ (lower right). The shaded regions are allowed by the COHERENT data for \cevens\!\!.}
	\label{fig:modscan}
\end{figure*}

\begin{figure*}[htb]
	\hspace*{-0.4cm} \includegraphics[scale = 0.21]{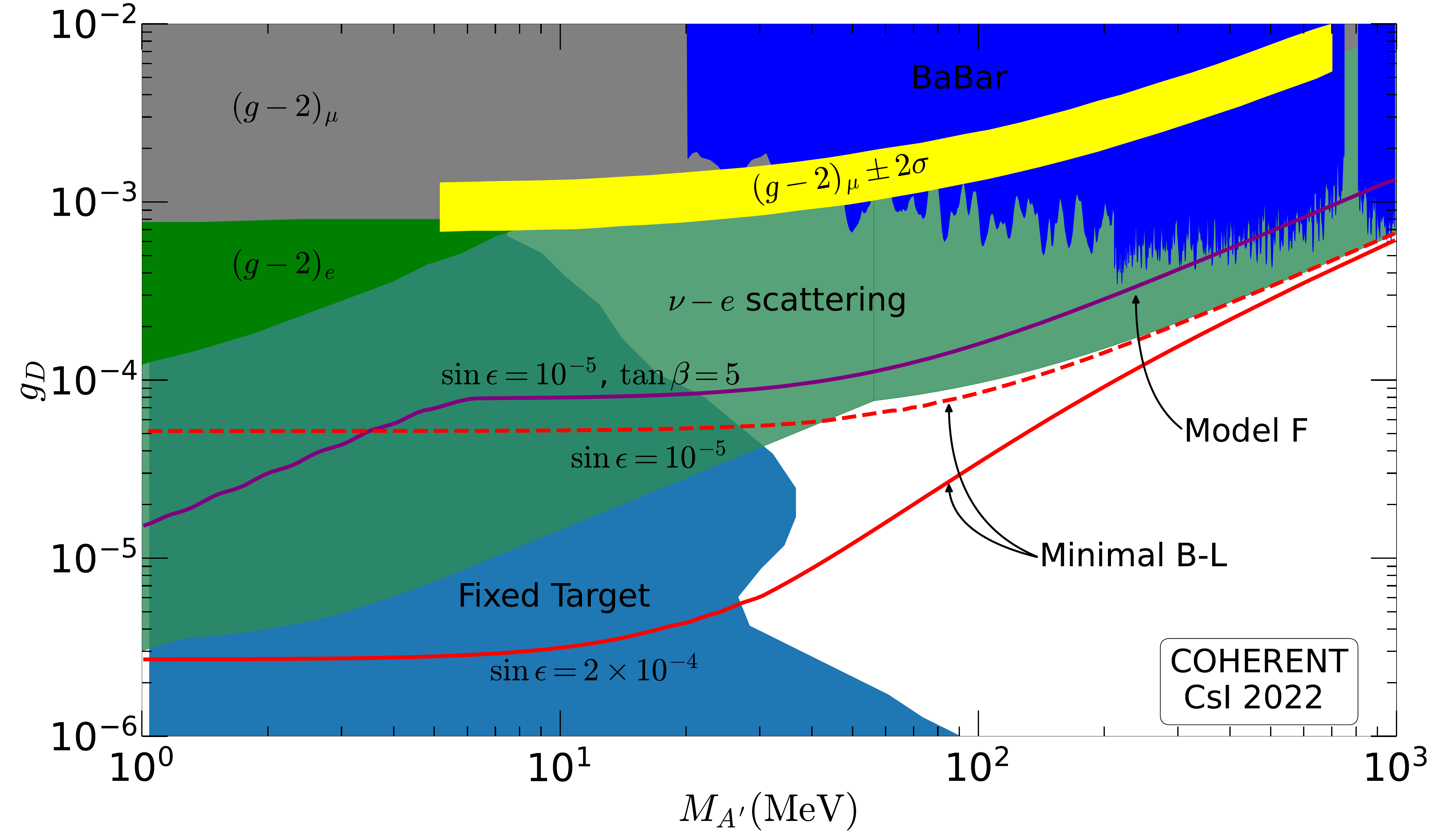} 
%	\vspace*{-1.4cm}
	\caption{\label{fig:litcomp}The exclusion plot of the bounds on the gauge coupling $g_D$ of the dark photon from laboratory experiments in the 1 MeV-1 GeV dark photon mass region, adapted from Ref.~\cite{PhysRevD.92.033009} with combined limits from our COHERENT data analysis, shown only for the two most promising scenarios.  
	}
\end{figure*}

Figure~\ref{fig:modcomparison} compares the exclusion regions of all the models considered so far in the $(g_D,\sin\epsilon)$ parameter space (the left panel) and in $(g_D,M_{A'})$ (the right panel). The analysis comparing the models  has been performed for an intermediate dark photon mass ($M_{A'}=50$ MeV), the parameter\footnote{It is observed that the choice of a value for the parameter $\tan\beta$ does not play a significant role for the most part of the $\tan\beta$ interval and indeed the sensitivity entirely disappears for $\tan\beta \gtrsim 5$.} $\tan\beta=5$, and for the CsI target, energy-dependent QF, and single bin assumption. From the graph on the left panel, the minimal $B-L$ model seems to give the most stringent bound in the entire region under the chosen circumstances while Models F and C are the second best such that for smaller values of the kinetic mixing parameter ($\sin\epsilon \lesssim 5\times 10^{-4}$)  Model F does better than Model C and vice versa in the larger mixing region. The general behaviors of the models as a function of $M_{A'}$ in the MeV--GeV range can be seen in the right panel where the minimal $B-L$ model has the most stringent exclusion boundary for $M_{A'}\gtrsim$ few MeV while Model F is the most sensitive one for the eV--keV dark photon mass.

After having discussed the effect of the COHERENT data for the \cevens experiment on the  parameter spaces of various representative two-Higgs doublet dark models as well as the minimal $B-L$ model, one may wonder what happens by varying the free dark charges $Q_u^\prime$ and  $Q_d^\prime$ which are different from the sets already listed in Table \ref{ChargeTable}. The results are depicted in Fig.~\ref{fig:modscan}. In the upper left graph, the light blue shaded region  is allowed $90\%$ C.L. by the COHERENT data in the  $(Q_u^\prime,Q_d^\prime)$ parameter space. The wider shaded strip represents the allowed region by the dark photon-mass constraint. The models are also marked on the graph and, for example, for the chosen inputs which are also indicated on the graph, Models E and F are ruled out by the data. The sensitivity of the allowed region on $\tan\beta$ is shown on the upper right graph for $\tan\beta=2$ and  $\tan\beta=20$ and the allowed region is slightly wider for  $\tan\beta=2$. In the lower row, the allowed regions are indicated for various $g_D$ values on the left panel and for various $M_{A'}$ values on the right panel. The size of the allowed region has effected significantly by varying $g_D$, which is somehow  less pronounced for the variation of the dark photon mass.

	Last but not least, in Fig.~\ref{fig:litcomp}, the exclusion regions from the neutrino-neucleus scattering  data of COHERENT Collaboration for the two most promising scenarios, the Model F and Minimal $B-L$ model, are displayed with other laboratory bounds relevant in the MeV--GeV range.  The figure is adapted from Fig. 6 in Ref.~\cite{PhysRevD.92.033009}  where the details of the laboratory experiments are given (see Table III there). As seen from Fig.~\ref{fig:litcomp} that there is a new region in 30 MeV--1 GeV range which is now probed and excluded by the COHERENT data depending on the value of the kinetic mixing parameter.

\section{Conclusion}
\label{conc}
More  than four decades after its first proposal, coherent elastic neutrino nucleus scattering was successfully measured by the COHERENT Collaboration in 2017, which has initiated a vast number of phenomenological and theoretical studies since then including physics beyond the SM. \cevens measurement can be used as a low-energy probe for the new physics searches and hidden sector coupled to the SM through portals is one of such scenarios. In the meantime, the discovery of the Higgs boson motivates the so-called 2HDMs where the mass of neutrinos could also be explained. The vector portal which is originally defined to  describe the interactions of the dark photon to the SM currents can be extended by taking the 2HDM as the visible sector instead of the SM. 

After briefly explaining the theoretical framework and listing seven representative models in Table \ref{ChargeTable} where the scalar doublets are allowed to have nonzero dark charges under the gauge group $U(1)_D$, we go on to provide analytical expressions for the differential cross sections for \cevens both in the SM and in the 2HDM extended with $U(1)_D$. The analyses of the COHERENT data for \cevens from the year 2017 to 2022 have been modified in various ways and some of these details including statistical analysis have been explained together with our basics to carry out the numerical study.

In this study we aim to find out the constraints on the parameter space of the 2HDMs extended with $U(1)_D$ by using the COHERENT neutrino scattering data. This has been achieved by looking at the effects of different factors and parameters which might play some role in the analysis.  These are the dark photon mass $M_{A'}$, kinetic mixing parameter $\sin\epsilon$, the dark gauge coupling $g_D$, the ratio of the VEVs $\tan\beta$ and the free dark charges $Q_u^\prime$, $Q_d^\prime$ as far as the theory is concerned. On the experimental side, there are additional factors like the number of energy and time bins used, data taken at different times for different target samples, which we call CsI 2017, CsI 2022, LAr A and LAr B, various quenching factor parametrizations. We also make a scan over the chosen models which are obtained by choosing values for $(Q_u^\prime$, $Q_d^\prime)$ as well as obtaining the allowed region by the COHERENT data on the $(Q_u^\prime$, $Q_d^\prime)$ parameters space in a much wider range.

The models show sensitivity to the mass of the dark photon, $M_{A^\prime}$ especially in the MeV-GeV range and this behavior disappears for the so-called minimal $B-L$ model for around $M_{A^\prime}\lesssim 1$MeV where the exact value depends on the kinetic mixing. There is a distinct difference in the behavior of the other models (Model C-F and B-L of the extended 2HDM) as $M_{A^\prime}\lesssim 1$MeV where the curves start showing a strong dependence on the value of $g_D$. Indeed, in the 2HDMs extended with $U(1)_D$, the dark photon mass receives contributions from the scalar sector being proportional to the parameter $g_D$. No matter how small the kinetic mixing is, there exists $g_D$ proportional contribution to the mass in addition to the $m_X$ term which brings the behavior at low dark photon mass tail (see for example the right panel of Fig.~\ref{fig:modcomparison}). Therefore  the minimal $B-L$ model gives the best bound for  $M_{A^\prime}\ge 1$MeV while the other models do much better for the lighter dark photon region. This might be taken as a way to distinguish them. It is also observed that the best bounds are obtained for the single bin case (1PE-1t), for the CsI 2022 data and for the QF taken to be constant (even though energy-dependent QF has been used throughout the numerical analysis as proposed by the COHERENT Collaboration in their latest analysis \cite{COHERENT:2021xmm}). In a set of plots, Fig.~\ref{fig:modscan}, the allowed bands on the $(Q_u^\prime$, $Q_d^\prime)$ plane have been shown and this could be used as a reference for better assessment of the scenarios beyond the chosen representative ones listed in Table \ref{ChargeTable}. In a final plot, our results are overlaid on a global picture and  a new region could further be excluded in the  30 MeV--1GeV dark-photon mass range by the COHERENT data depending on the value of the kinetic mixing parameter. Further data available at low energies may either probe new physics scenarios like 2HDM better or even point out a smoking gun signal which may shape out the physics beyond the SM.

%\begin{acknowledgements}
\acknowledgments
	This work is supported in part by the Scientific and Technological Research Council of Turkey (TUBITAK) grant 118F390.
%\end{acknowledgements}

%\clearpage
%\bibliographystyle{unsrt}
%\bibliographystyle{myunsrt}
%\bibliographystyle{apacite}
\bibliography{mycevensV2}

\end{document}